\documentclass[electronic]{vgtc}             

\ifpdf
  \pdfoutput=1\relax                   
  \usepackage{graphicx}                
  \DeclareGraphicsExtensions{.pdf,.png,.jpg,.jpeg} 
\else
  \usepackage{graphicx}                
  \DeclareGraphicsExtensions{.eps}     
\fi%

\graphicspath{{figures/}{pictures/}{images/}{./}} 

\usepackage{microtype}                 
\PassOptionsToPackage{warn}{textcomp}  
\usepackage{textcomp}                  
\usepackage{mathptmx}                  
\usepackage{times}                     
\usepackage{cite}                      
\usepackage{booktabs}                  

\usepackage{bookmark}
\usepackage{subcaption}
\usepackage{array}
\usepackage{multirow}
\usepackage{xcolor,colortbl}

\definecolor{Gray}{gray}{0.85}
\definecolor{LightCyan}{rgb}{0.88,1,1}
\definecolor{LightRed}{rgb}{1,0.88,1}

\newcolumntype{a}{>{\columncolor{Gray}}c}
\newcolumntype{b}{>{\columncolor{LightCyan}}c}
\newcolumntype{d}{>{\columncolor{LightRed}}c}

\usepackage{amsmath}
\usepackage{amsfonts}
\usepackage{amstext}

\onlineid{1015}

\title{Scalable In Situ Lagrangian Flow Map Extraction: Demonstrating the Viability of a Communication-Free Model}

\author{Sudhanshu Sane\thanks{e-mail: ssane@uoregon.edu}\\ %
        \scriptsize University of Oregon %
\and Abhishek Yenpure\\ %
     \scriptsize University of Oregon %
\and Roxana Bujack\\ %
     \scriptsize Los Alamos Nat'l Laboratory %
\and Matthew Larsen\\ %
     \scriptsize Lawrence Livermore Nat'l Laboratory %
\and Kenneth Moreland\\ %
     \scriptsize Sandia Nat'l Laboratories %
\and Christoph Garth\\ %
     \scriptsize Technische Universit{\"a}t Kaiserslautern %
\and Hank Childs\\ %
     \scriptsize University of Oregon}

\abstract{
We introduce and evaluate a new algorithm for the in situ extraction of Lagrangian flow maps, which we call
Boundary Termination Optimization~(BTO).
Our approach is a communication-free model, requiring no message passing or synchronization between processes, improving scalability, thereby reducing overall execution time and alleviating the encumbrance placed on simulation codes from in situ processing.
We terminate particle integration at node boundaries and store only a subset of the flow map that would have been extracted by communicating particles across nodes, thus introducing an accuracy-performance tradeoff.
We run experiments with as many as 2048 GPUs and with multiple simulation data sets.
For the experiment configurations we consider, our findings demonstrate that our communication-free technique saves as much as 2x to 4x in execution time in situ, while staying nearly as accurate quantitatively and qualitatively as previous work.
Most significantly, this study establishes the viability of approaching in situ Lagrangian flow map extraction using communication-free models in the future. 

} 

\begin{document}


\firstsection{Introduction}
\maketitle
Over the last decade, the compute capabilities of supercomputers 
have increased by a whole order of magnitude relative to the I/O 
capabilities of these machines.
As a result, computational simulations on supercomputers are now able
to generate data much faster than they can store it.
Although temporal subsampling, e.g., saving every $N^{th}$ time step for subsequent \textit{post hoc}
visualization, has always been practiced, the inherent I/O bottleneck on modern supercomputers is forcing this subsampling to become severe.
Such sparse temporal settings can hinder accurate reconstruction and 
exploration by domain scientists.
\textit{In situ} processing~\cite{bauer2016situ},
i.e., coupling analysis routines with the simulation code and processing data as it is generated,
addresses this issue by limiting I/O.
Additionally, access to all spatio-temporal data creates opportunities
that were not available when storing only a fraction of 
time slices to disk.
In situ processing has gained significant momentum 
in the last half-decade, including myriad research efforts, devoted workshops (ISAV, WOIV), and a recent Dagstuhl seminar~\cite{Dagstuhl19}.


A key issue for in situ processing, particularly in scientific exploration use cases, is the lack of a priori knowledge regarding the precise desired analysis and visualizations.
If a priori knowledge exists, then in situ analysis and visualization is a matter of integrating the desired routines into a simulation code. 
If not, then in situ visualization is more complicated, since it is unclear what activities to do.
A common solution to this problem is to transform the data to a reduced form, small enough that it can be stored to disk for post hoc exploration.
We refer to this paradigm as ISR-PHE
(in situ reduction, post hoc exploration).
Typically, the ISR-PHE paradigm involves navigating a tradeoff between data integrity and data reduction, and
ensuring that the corresponding routines operate within in situ constraints.
%
%
In this paper, we consider time-varying flow visualization in the context of ISR-PHE. 

Previous work by Agranovsky et al.~\cite{agranovsky2014improved}
has shown that ISR-PHE of time-varying vector field data via a Lagrangian representation 
is a compelling approach.
During the in situ phase, particles trajectories are integrated using the simulation generated vector field. 
These trajectories, referred to as \textit{basis flows}, encode the underlying time-varying behavior of the simulation flow field and are stored to disk. 
In the post hoc phase, new trajectories can be interpolated from the
basis flows, enabling exploration.

Accurate particle advection for a time-varying flow field requires having
access to complete spatio-temporal data.
The traditional (Eulerian) approach is handicapped
by having access to only a sparse temporal subsampling, 
leading to flow visualizations containing high numerical approximation errors.
In contrast, the ISR-PHE Lagrangian approach
does have full access to complete spatio-temporal data in situ, allowing it to calculate its basis flows with
complete accuracy.
As a result, somewhat surprisingly, 
this approach does not require a tradeoff between data integrity and data reduction ---
it improves data integrity 
even while achieving data reduction compared to the traditional approach 
(saving mesh-based vector field data at regular intervals).
The Agranovsky et al. method demonstrated a 10x improvement
in accuracy using the same storage as the traditional approach,
and a 1.2x improvement in accuracy with a 64x reduction in storage.
%
%
%
%
%
%
%
%
%
Achieving these significantly improved accuracy-storage propositions, however, involves expensive in situ routines, particularly at scale.
With our work, we build on the existing excellent accuracy-storage propositions, and propose an accuracy-performance tradeoff that allows Lagrangian analysis to operate within in situ constraints while maintaining nearly as accurate time-varying vector field exploration capabilities.
%
%

A major concern for in situ processing is the encumbrance placed on the simulation code.
In situ analysis tasks are allocated a limited resource budget, and,
%
%
since analysis is coupled with a simulation, analysis performance directly impacts the overall performance.
%
%
In situ calculation of a Lagrangian representation, however, involves distributed memory particle advection to calculate sets of basis flows (pathlines), introducing a heavy communication overhead.
Although addressing the scalability of distributed memory integral curve computation has been extensively studied in a post hoc setting, current techniques often do not operate within in situ constraints.
%
Further, communicating particle information between nodes every cycle is expensive and will only become more so as supercomputers get larger.
%
To address the challenge of in situ Lagrangian analysis scalability, we evaluate the viability of a communication-free model.
In the context of distributed memory integral curve computation, this is a relatively unexplored idea.
However, as opposed to post hoc flow visualization (toward which most research efforts have been directed), we believe our objective of extracting a Lagrangian flow map in situ permits such a model.

In this paper, we propose the Boundary Termination Optimization (BTO), a simple, yet novel, communication-free algorithm to improve the performance of in situ \textit{extraction} of Lagrangian flow maps.
%
%
%
%
%
We demonstrate the viability and limitations of this method by considering multiple simulation data sets and parameter configurations.
Our study finds that a communication-free model, for the several practical configurations considered, achieves 3x speed-up while calculating flow maps in situ and can approximate the complete flow map while maintaining over 96\% reconstruction accuracy in several cases compared to using communication within Lagrangian analysis.
%

\section{Background and Related Work}
A key operation for calculating a Lagrangian flow map in situ is computing pathlines in a distributed memory environment.
We first discuss relevant Lagrangian analysis works, followed by a brief categorization of distributed memory techniques and the constraints preventing existing approaches from being adopted in situ.
We end with a description of the steps involved in calculating a Lagrangian flow map in situ.

\subsection{Lagrangian Analysis}
In the Lagrangian specification, flow field information is stored using integral curves, where each integral curve encodes the trajectory of a single massless particle and describes its properties as it travels through space.
Each integral curve provides insight regarding flow behavior in the vicinity of the particle's trajectory~\cite{bujack2015lagrangian}.
Collectively, a large number of integral curves spanning the spatial domain can be defined in terms of a flow map, i.e., a Lagrangian representation of the flow field.
The flow map $F_{t_0}^{t}(x_0):\mathbb R^d \times \mathbb R \times \mathbb R  \to \mathbb R^d$ describes where a particle starting at position $x_0\in \mathbb R^d$ and time $t_0\in \mathbb R$ moves to in the time interval $[t_0,t]\subset \mathbb R$~\cite{garth2007efficient}.

Lagrangian methods have been increasingly used for flow visualization in the past decade.
Lagrangian coherent structures (LCS), introduced by Haller et al.~\cite{haller2001distinguished, haller2000lagrangian, haller2000finding}, is a popular technique to visualize attracting and repelling surfaces.
Given the high computational cost of calculating LCS, several efforts have been aimed at accelerating its computation and visualization~\cite{garth2007efficient,garth2009visualization,sadlo2007efficient,sadlo2011time}.
Further, LCS are visualized for varying application domains, such as large eddy simulations used to assess the transport properties of multiscale ocean flows~\cite{ozgokmen2012multi} and the design of energy-efficient buildings with respect to air circulation~\cite{schindler2012lagrangian}.

More recently, in situ Lagrangian analysis has been used to extract time-varying flow field information.
Most relevant to our work, Agranovsky et al.~\cite{agranovsky2014improved} extracted flow maps over a vector field in situ to represent the behavior of the flow for intervals of time.
We describe this work in more detail in Section~\ref{sec:lagrangian-mpi}.
Sane et al.~\cite{sane2018revisiting} extended the evaluation of Lagrangian in situ analysis by considering spatiotemporal tradeoffs across a range of configurations.
Working with an SPH~\cite{GM77} simulation, Chandler et al.~\cite{chandler2015interpolation} stored the calculated flow maps in a modified k-d tree and queried it for pathline interpolation.
However, none of these works considered the scalability of extracting flow maps in situ.

Several works have presented Lagrangian-based advection methods for post hoc flow reconstruction using Lagrangian flow maps as input. 
Hlawatsch et al.~\cite{hlawatsch2011hierarchical} employed a hierarchical scheme post hoc to decrease the number of integration steps by constructing longer pathlines using previously computed Lagrangian-based trajectories.
Agranovsky et al.~\cite{agranovsky2011extracting} optimized pathline interpolation using scattered particles and considered the use of Moving Least Squares and Barycentric coordinate interpolation.
Chandler et al.~\cite{chandler2016analysis} presented an analysis showing Lagrangian interpolation error correlates with divergence in flow fields.
Bujack et al.~\cite{bujack2015lagrangian} identified local truncation error propagation as a source of error in Lagrangian-based advection methods. 
Extending this work, Hummel et al.~\cite{hummel2016error} provided theoretical upper error bounds and pathline uncertainty visualizations.
Aiming to reduce the error propagation, Sane et al.~\cite{sane2019interpolation} proposed a pathline interpolation scheme that is capable of using longer particle trajectories varying in seed placement and duration.
In this work, we use Barycentric coordinate interpolation as our Lagrangian-based advection method for flow reconstruction~\cite{agranovsky2011extracting,agranovsky2014improved}.

While these studies have broadened and improved in situ Lagrangian analysis from an aesthetics, reconstruction accuracy, and storage perspective, very little work has been devoted to performance considerations and viability.
%
%

\subsection{Distributed Integral Curve Computation}
Efficient computation of integral curves in a distributed memory environment is an extensively researched field.
However, the majority of works are limited to steady state vector fields in a post hoc environment.
%
%
The primary challenge addressed by these works is improving the scalability, load balance, and overall efficiency of distributed memory integral curve computation. 
Typical parallelization strategies adopted to improve performance are parallelize-over-data~\cite{sujudi1996integration,peterka2011study,chen2008optimizing,yu2007parallel,nouanesengsy2011load},
parallelize-over-particles~\cite{dinan2009scalable,camp2011evaluating,muller2013distributed,guo2014advection},
or a hybrid approach~\cite{pugmire2009scalable,kendall2011simplified,lu2014scalable}.
Parallelize-over-data techniques determine a domain decomposition and assign a subset of the total domain to each node.
While calculating integral curves, these methods communicate particles between processors when required.
Parallelize-over-particles techniques assign a set of particles to each node and load data from disk on demand. 
For a complete review of the algorithms, we refer the reader to a recent survey by Zhang et al.~\cite{zhang2018survey}.

Although involving similar keywords as our own work, the following related works are only applicable to steady state vector fields in a post hoc setting.
%
%
%
Bleile et al.~\cite{bleile2017accelerating} accelerated streamline calculation by swapping traditional Eulerian and Lagrangian-based advection at node boundaries.
In this case, after a particle is communicated across a boundary, a previously computed mapping is used to transport the particle across the entire node. 
Liao et al.~\cite{Liao2019ScalablePF} presented a communication-free 3D LIC technique.
They limit communication by using a preprocessing step to regroup unstructured grid cells and restricting particle advection to within the confines of a single cell.
Existing algorithms are difficult to adopt in the context of in situ calculation of pathlines.
Operating in an in situ environment introduces new constraints with regard to which process can access what domain subset in space and time.
First, domain decomposition and distribution are simulation-determined.
Second, the time-varying nature of simulation data in conjunction with in situ memory constraints means only a single time step can be accessed at any given time, and communicating this information across processors each step is prohibitively expensive.
These constraints complicate techniques like data prefetching, rearrangement, or completing particle advection within a node before particle exchange. 
(In situ methods are currently limited to advancing a particle by a single step before the vector field changes.)
%
Further, the problem of poor scalability remains when considering a large number of processors and communication between them every cycle. 

\subsection{In Situ Extraction of Lagrangian Flow Maps}
\label{sec:lagrangian-mpi}
Agranovsky et al.~\cite{agranovsky2014improved} were the first to use a Lagrangian representation for ISR-PHE of a time-varying flow field. 
The method involves two distinct phases.
The first phase involves in situ extraction of sets of basis flows (i.e., pathlines) and the second phase involves flow reconstruction using a Lagrangian-based advection method.

In the in situ phase, particles are seeded along a uniform grid and advected for a predetermined interval of time.
In this context, an interval is defined as the number of cycles particles are advected for before saving their end positions and resetting them along a uniform grid.
Thus, sets of basis flows are calculated in batches for nonoverlapping intervals of time. 
The number of seeds placed is configurable and depends on the desired data reduction.
The data reduction is based on the vector field grid size, i.e., a 1:1 configuration uses as many particles as grid points, and a 1:27 configuration uses one particle for every twenty-seven grid points. 
Importantly, in the Agranovsky method, each process communicates particle exchange information after each simulation step, which contributes significantly to the total execution time.
On ``write cycles,'' i.e., the cycles where data is stored to disk, each compute node returns its particles to their originating nodes.

In the post hoc phase, new particle trajectories for a desired flow analysis can be generated by interpolating the extracted basis flows.
Tracing a new pathline begins with identifying the right neighborhood of basis flows to ``follow'' for an interval of time.
Successive sets of basis flows can then be used to stitch together a complete pathline.
This Lagrangian-based advection scheme enables scientists to explore the flow field, considering particle trajectories that were not saved as basis flows.
%


\section{Boundary Termination Optimization}
\label{sec:method}
In this section, we first define requirements for extracting flow field information as a set of pathlines in situ.
Next, we describe Boundary Termination Optimization~(BTO), which results in the communication-free technique Lagrangian-BTO and our demonstration of it's viability is the main contribution of this study.
Further, we discuss the accuracy-performance tradeoff when comparing a communication-based versus communication-free model. 
Finally, we provide the details of the Lagrangian-based advection scheme we use for reconstruction and a theoretical error analysis.

The requirements surrounding the calculation of integral curves in distributed memory varies depending 
on whether the computation is for the purpose of post hoc visualization and analysis or in situ extraction of a flow map. 
%
%
Whereas in a post hoc setting an integral curve must continue particle integration across node boundaries, this is not necessary to calculate a Lagrangian representation of the flow field.
We identify different requirements when extracting a Lagrangian flow map in situ and define them as follows:
\begin{enumerate} 
\item Extraction of a flow map or set of pathlines in situ should demonstrate good scalability. 
\item Flow field reconstruction using extracted pathlines should be accurate.
\end{enumerate}

The method described in Section~\ref{sec:lagrangian-mpi} demonstrated only the second requirement. 
We implement the method by Agranovsky et al.~\cite{agranovsky2014improved} using MPI\cite{gropp1996high} for communication and refer to our implementation of this technique as Lagrangian-MPI.
In this paper, we use Lagrangian-MPI as a baseline for comparison.
%

\subsection{In Situ Extraction Using BTO}
Our contribution with this work is a simple communication-free algorithm for extracting a Lagrangian flow map in situ.
%
The benefit of this approach is that it has less execution time and improved scalability characteristics, which reduces the burden on the simulation code.
To improve performance, our approach requires a small modification to Lagrangian-MPI: eliminate information exchange and synchronization. 

%
%

\begin{figure}[t]
\begin{subfigure}{0.495\linewidth}
\centering
\includegraphics[width=0.6\linewidth]{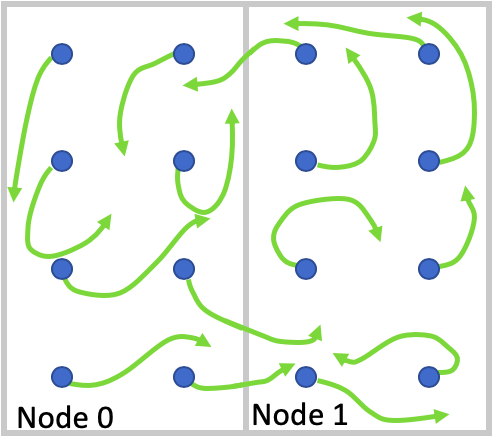}
\caption{Lagrangian-MPI}
\label{notional_insitu_mpi}
\end{subfigure}
\hfill
\begin{subfigure}{0.495\linewidth}
\centering
\includegraphics[width=0.6\linewidth]{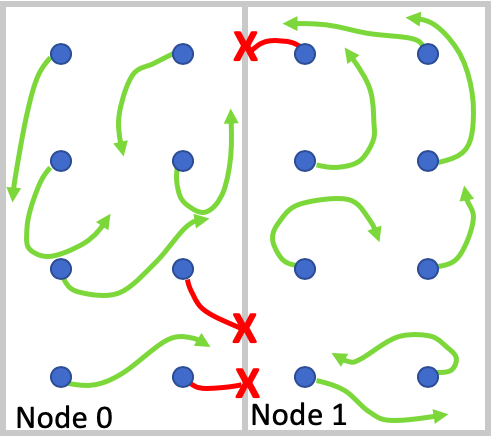}
\caption{Lagrangian-BTO}
\label{notional_insitu_bto}
\end{subfigure}
\caption{Notional examples of calculating basis flows. Only green trajectories are stored to disk.} 
\label{basisflows}
\end{figure}

%
Similar to Lagrangian-MPI, we use an initially uniform seed placement and advect particles for predetermined nonoverlapping intervals of time.
However, as opposed to continuing particle integration across node boundaries, our approach terminates and discards these particle trajectories.
Figures~\ref{notional_insitu_mpi} and~\ref{notional_insitu_bto} illustrate notional examples of basis flows calculated by Lagrangian-MPI and Lagrangian-BTO, respectively.
Thus, we store only those particle trajectories that remain within the domain until the end of the interval.
Terminating particles that require communication across node boundaries to continue trajectory integration, allows the approach to remain communication-free.
%
Since processors do not exchange particles, 
the Lagrangian analysis operator on each processor can operate independently and asynchronously. 
%
%

We build both Lagrangian analysis techniques, i.e., Lagrangian-MPI and Lagrangian-BTO, as in situ analysis filters using the VTK-m~\cite{moreland2016vtk}, VTK-h~\cite{larsen2017alpine}, and Ascent~\cite{larsen2017alpine} libraries. 
VTK-m is a platform portable scientific visualization library for shared memory parallel environments.
VTK-h is a distributed memory wrapper around VTK-m. 
Ascent is an in situ visualization infrastructure that we use to both integrate with simulations and create a workflow when loading data sets from disk.
%

The Lagrangian-BTO filter has two operations to perform: particle advection and particle management. 
Particle advection is performed using RK4 interpolation implemented as a VTK-m worklet~\cite{pugmire2018performance}. 
%
%
Particle management involves tracking particle trajectories, evaluating the validity, and managing memory to prevent invalid particles from being launched on GPU threads during advection.
The Lagrangian-MPI filter has to perform three operations: particle advection, particle management, and communication.
Consistent with the Agranovsky approach, communication of particles and particle information between ranks is performed using asynchronous, non-blocking, buffered MPI communication.
Additionally, particle management further includes tracking of \textit{internal} and \textit{external} particles in order to return particles to originating nodes at write cycles.

\textbf{Accuracy-Performance Tradeoff ---} The loss of information in the form of terminated and discarded particle trajectories reduces the quality of flow reconstruction.
Flow information that will typically be lost near boundaries can, however, be interpolated using additional information from adjacent processes post hoc.
With respect to simulation overhead, the communication-free model offers a reduced cost and improved scalability.
We believe evaluating the accuracy-performance tradeoff and determining the viability of a communication-free model as an alternative low-cost choice for a scientist or researcher is valuable.
Our hypothesis regarding this method is that the execution time will be substantially improved (since there is no communication required), but the accuracy will be only modestly affected.
%

%
%
Finally, the Agranovsky work demonstrated that the Lagrangian method can be much more
accurate than the Eulerian approach, including some cases where the accuracy improved by over 10x.
%
Even though our practice of terminating particles at the boundary will reduce accuracy
compared to the Agranovsky approach, we still would be much more accurate than
the Eulerian approach.
%
%
If we also can demonstrate significantly faster execution times, then we
believe our proposed method would be appealing to future researchers and domain scientists in
many settings.
%

\begin{figure}[t]
\begin{subfigure}{0.495\linewidth}
\centering
\includegraphics[width=0.6\linewidth]{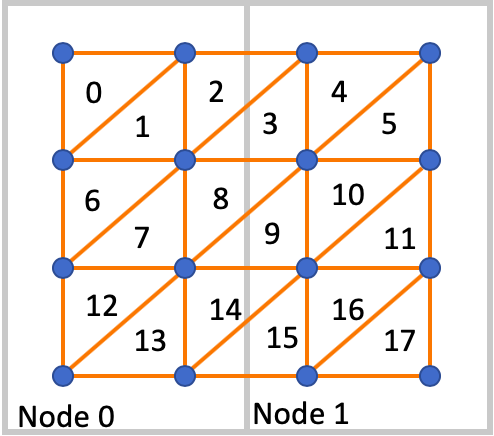}
\caption{Neighborhoods using~\ref{notional_insitu_mpi}}
\label{notional_posthoc_mpi}
\end{subfigure}
\hfill
\begin{subfigure}{0.495\linewidth}
\centering
\includegraphics[width=0.6\linewidth]{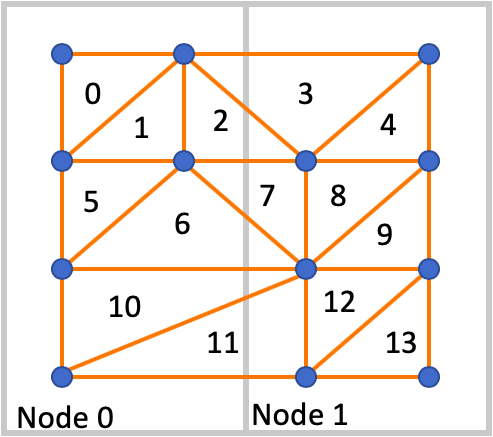}
\caption{Neighborhoods using~\ref{notional_insitu_bto}}
\label{notional_posthoc_bto}
\end{subfigure}
\caption{Notional examples of neighborhoods produced using triangulation over basis flows shown in Figure~\ref{basisflows}. The blue circles represent starting locations of basis flows saved to disk.}
\label{triangulation}
\end{figure}

\subsection{Post Hoc Reconstruction}
\label{sec:reconstruction}
To reconstruct the complete flow map or to trace new pathlines, we set up a parallelize-over-data distributed memory Lagrangian-based advection scheme that is conceptually similar to previous work~\cite{agranovsky2014improved,chandler2015interpolation,sane2018revisiting}.
For a given interval, each process loads basis flows generated from its own and adjacent processes for that interval.
%
%
For any new trajectory to be calculated using basis flows, the particle must identify a neighborhood (convex hull) of basis flows to follow for the duration of an interval.
%
%
We perform a Delaunay triangulation over the start locations of loaded basis flows to identify neighborhoods for tracing new particle trajectories.
%
%
Figure~\ref{triangulation} illustrates a notional example of triangulation using basis flows from the two nodes demonstrated in Figure~\ref{basisflows}.
Using basis flows data from adjacent neighboring processes, i.e., flow information from all directions, allows approximation of regions where information has been lost. 
Once a particle neighborhood is identified, the location of the particle at the end of the interval can be calculated by following the neighborhood of basis flows.
In our work, we use Barycentric coordinate interpolation to calculate the particle's end position.
To calculate a pathline for a duration greater than the interval of basis flows, a trajectory can be stitched together by using basis flows of successive nonoverlapping intervals.

%
%
%
We implemented our reconstruction workflow using CGAL for 3D parallel Delaunay triangulation~\cite{fabri2011cgal} and the vtkProbeFilter from the VTK library~\cite{schroeder2004visualization} for Barycentric coordinate interpolation.
Further, we performed all our reconstruction and accuracy measurements on our in-house research cluster.

\subsection{Theoretical Error Analysis}
We use a one-dimensional linear interpolation $L$ of a function $f:\mathbb R\to \mathbb R$ for $x\in[x_0,x_1]\subset\mathbb R$
\begin{footnotesize}
\begin{eqnarray}
\begin{aligned}\label{linearInterpolation}
L_{f(x_0),f(x_1)}(x) = \frac{x-x_0}{x_1-x_0}f(x_1) + \frac{x_1-x}{x_1-x_0}f(x_0).
\end{aligned}
\end{eqnarray}
\end{footnotesize}
The higher dimensional result satisfies Equation~\eqref{linearInterpolation} for each component.

In our approach, a particle starting at $x_1$ that reaches the node boundary during the interval of advection is terminated.
Thus, its function value (the flow map) is not known. 
However, we can reconstruct it from its known neighbors $L_{f(x_0),f(x_2)}(x_1)$. 
Consider a particle $x\in[x_0,x_1]\subset\mathbb R$ whose path is interpolated post hoc using this reconstructed value. 
We get the same result as if we had used the closest existing neighbors directly, because let $L'$ denote $L_{f(x_0),L_{f(x_0),f(x_2)}(x_1)}(x)$, then
\begin{footnotesize}
\begin{eqnarray}
\begin{aligned}\label{doubleInterpolation}
L' &\overset{\eqref{linearInterpolation}}= \frac{x-x_0}{x_1-x_0}L_{f(x_0),f(x_2)}(x_1) + \frac{x_1-x}{x_1-x_0}f(x_0)\\
&\overset{\eqref{linearInterpolation}}= \frac{x-x_0}{x_1-x_0}(\frac{x_1-x_0}{x_2-x_0}f(x_2) + \frac{x_2-x_1}{x_2-x_0}f(x_0)) + \frac{x_1-x}{x_1-x_0}f(x_0)\\
&= \frac{x-x_0}{x_2-x_0}f(x_2) + \frac{(x_2-x_1)(x-x_0)+(x_2-x_0)(x_1-x)}{(x_2-x_0)(x_1-x_0)} f(x_0)\\
&= \frac{x-x_0}{x_2-x_0}f(x_2) + \frac{x_2-x}{x_2-x_0}f(x_0)\\
&\overset{\eqref{linearInterpolation}}= L_{f(x_0),f(x_2)}(x).
\end{aligned}
\end{eqnarray}
\end{footnotesize}
Bujack et al.~\cite{bujack2015lagrangian} previously established that the post hoc interpolation 
of pathlines is a numerical one-step integration method~\cite{GH10}. 
Its accuracy is bounded by its global truncation error at stitching step $n\in\mathbb N$ of 
\begin{footnotesize}
\begin{equation}\begin{aligned}\label{global}
e_{n}\leq\frac {d^2}8(t_n-t_0) h_x^2\max_{\tau\in[t_0,t_n]}\max_{\zeta\in\mathbb R^d}
\| H_{{\dot F}_{t_{j-1}}^{\tau}}(\zeta)\|_{\infty}e^{L(t_n-t_0)}
\end{aligned}\end{equation}
\end{footnotesize}
with dimension $d$, start time $t_0$, end time $t_n$, spatial Lipschitz constant $L$, Hessian $H$ of the temporal derivative of the flow map $\dot F$, and spatial distance $h_x$ between the basis flows.
As a result, the interpolation error is $O(h_x^2)$ if the flow map has bounded second derivatives in space and first derivatives in time, which is a reasonable assumption for a differentiable vector field, because the solutions of an initial value problem depend smoothly on the initial conditions and time~\cite{hartman1973ordinary}.

Equation~\eqref{doubleInterpolation} shows that the error bound $O(\tilde h_x^2)$ still holds but with a larger $\tilde h_x>h_x$. 
Its size is determined by the size of missing information, which is limited by the maximum distance particles can move in one time interval.
If a particle is seeded further than $\max_{x_i\in\mathbb R^d} \max_{j=1..n} \| F_{j-1}^j(x_i)\|$ away from the node boundary, it cannot reach it and must therefore have the correct flow map information.
In the worst case, we can have missing information on both sides of the boundary, and therefore we get
\begin{footnotesize}
\begin{eqnarray}
\begin{aligned}\label{hx}
\tilde h_x&\leq 2h_x+2\max_{x_i\in\mathbb R^d} \max_{j=1..n} \| F_{j-1}^j(x_i)\|\\
&\leq 2h_x+2 h_t \max_{x\in\mathbb R^d} \max_{t\in[t_{j-1},t_j]}\| v(x,t)\|
\end{aligned}
\end{eqnarray}
\end{footnotesize}
with the underlying velocity field $v:\mathbb R^d \times \mathbb R \to \mathbb R^d$ and the temporal step size $h_t$, which is the interval time between storing data to disk, which is usually around one-thousandth of the total integration time.
Please note that the future increase of the global truncation error of a particle that traverses this region can continue even after it has left the region, but for all particles that never enter the region close to the boundary, the original error bound holds.
%

%
\section{Study Overview}
\label{sec:evaluation}
In this section, we describe our experiments to evaluate performance and reconstruction accuracy, and the metrics we use to measure the same. 

\subsection{Experiment Setup} 
To evaluate the viability of the Lagrangian-BTO analysis filters, we set up two workflows.
\begin{itemize}
\item \textbf{WF1:} In situ weak scaling study to evaluate speed-up.
\item \textbf{WF2:} Evaluation of reconstruction accuracy by varying parameters used in flow map generation. 
\end{itemize}

For the WF1 workflow, Ascent is directly connected to a simulation code.
Lagrangian analysis parameters are specified as input to the Ascent pipeline.
The simulation code generates vector field data and pauses when it invokes Ascent calls in order to perform in situ analysis.
We scaled the number of processors used and proportionately increased the size of the simulation grid.
Our \textbf{WF1} experiments provide insight into the performance of both Lagrangian analysis filters at scale.

For the WF2 workflow, we load data set files from disk to create a theoretical in situ setup, and then call Ascent to initiate the Lagrangian analysis filters.
We consider a fixed number of MPI tasks and nodes for \textbf{WF2}, i.e., we use a fixed domain decomposition for all tests.
Our \textbf{WF2} experiment configurations are designed to understand the range of specifications under which Lagrangian-BTO extracts accurate flow maps and to identify the potential limitations of using this method.
We vary the value of the interval parameter (i.e., the number of cycles we wait before writing to disk) to understand the effect of longer advection intervals on flow map generation using the Lagrangian-BTO approach.
In general, the longer the interval, the greater the probability of particles reaching the boundary and being terminated.
Further, we consider the 
impact of various data reduction options.
When using a very sparse number of particles
to capture the behavior of the flow, the effect of losing particles to boundary termination could be greater.
As mentioned in Section~\ref{sec:lagrangian-mpi}, data reduction values are described in the form of 1:X, i.e., one particle for every X grid points.

\subsection{Performance Metric} 
We measure performance in terms of the execution time of the Lagrangian analysis filters.
All of our timings are measured using the timing functionality in Ascent.
We measure the average time per cycle, which includes time to perform particle advection, particle management, and also communication in the case of Lagrangian-MPI. 
To simplify understanding the scalability of both Lagrangian analysis filters with respect to communication costs, we exclude cycles at the end of an interval.
These cycles include a communication cost incurred to return all particles to the respective origin nodes for Lagrangian-MPI and an I/O cost to write information to disk for both methods.
In this paper, we do not analyze the parallel I/O times, and we consider I/O optimization methods to be beyond the scope of this work. 

\subsection{Accuracy Metric} 
\label{accuracy_metric}
We measure the accuracy of flow field reconstruction by the Lagrangian-BTO analysis filter relative to the accuracy achieved by the corresponding Lagrangian-MPI analysis filter.
Comparisons of Lagrangian-MPI to the traditional Eulerian method can be found in previous works~\cite{agranovsky2014improved,sane2018revisiting}.
We measure total flow volume error using the average L2-norm over all samples considered.
The total average L2-norm is calculated as
\begin{equation}
        \frac{1}{p} \sum_{i=0}^{p} \lvert \lvert b_{i,t}  -  m_{i,t}\rvert \rvert
\end{equation}
where $p$ is the total number of particles, $b_{i,t}$ is the location of a Lagrangian-BTO interpolated particle $i$ at time $t$, and $m_{i,t}$ is the location of the Lagrangian-MPI particle $i$ at time $t$.

For a given total average L2-norm value \textbf{L}, the reconstruction accuracy percentage is proportional to the length of cell side \textbf{C} for that specific configuration, and is calculated as

\begin{equation}
Accuracy\%=\frac{C-L}{C}\times100
\end{equation}

We note that we use the total average L2-norm in two contexts. 
First, to measure error when reconstructing the complete flow map as generated by the Lagrangian-MPI method.
Second, to measure error of new pathlines traced using basis flows generated by both methods when compared to a ground truth.

In addition to the above total flow volume error measure, we report the maximum L2-norm in two forms ---  the greatest maximum L2-norm across all interval reconstructions, i.e., the error of the least accurately interpolated single basis flow, and the average maximum L2-norm across all intervals.

\subsection{Data Sets} 
We consider four data sets, namely, the Cloverleaf3D simulation, Arnold-Beltrami-Childress flow, Jet flow simulation, and Nyx cosmology simulation.

%

\subsection{Runtime Environment}
We tested the Lagrangian analysis techniques by running our experiments on Summit (supercomputer at ORNL).
Each node of Summit has two IBM Power9 CPUs, each with 22 cores running at 3.8 GHz and 512 GBytes of DDR4 memory.
Further, nodes on Summit have enhanced on-chip acceleration with each CPU connected via NVLink to 3 GPUs, for a total of 6 GPUs per node.
Each GPU is an NVIDIA Tesla V100 with 5120 CUDA cores, 6.1 TeraFLOPS of double precision performance, and 16 GBytes of HBM2 memory.

\section{Results}
We organize our results into four subsections,~\ref{sec:clover} to~\ref{sec:nyx}.
Each subsection is focused on one data set.
Specifically, in subsection~\ref{sec:clover}, we consider the Cloverleaf3D simulation with workflow \textbf{WF1} for our weak scaling study and workflow \textbf{WF2} for our strong scaling study.
In subsections~\ref{sec:abc},~\ref{sec:jet4}, and~\ref{sec:nyx}, we consider the ABC, Nyx, and Jet flow simulations, respectively, and run experiments with workflow \textbf{WF2}.

\begin{figure*}[!t]
\begin{subfigure}{0.245\textwidth}
\raggedright
\includegraphics[width=1.05\linewidth]{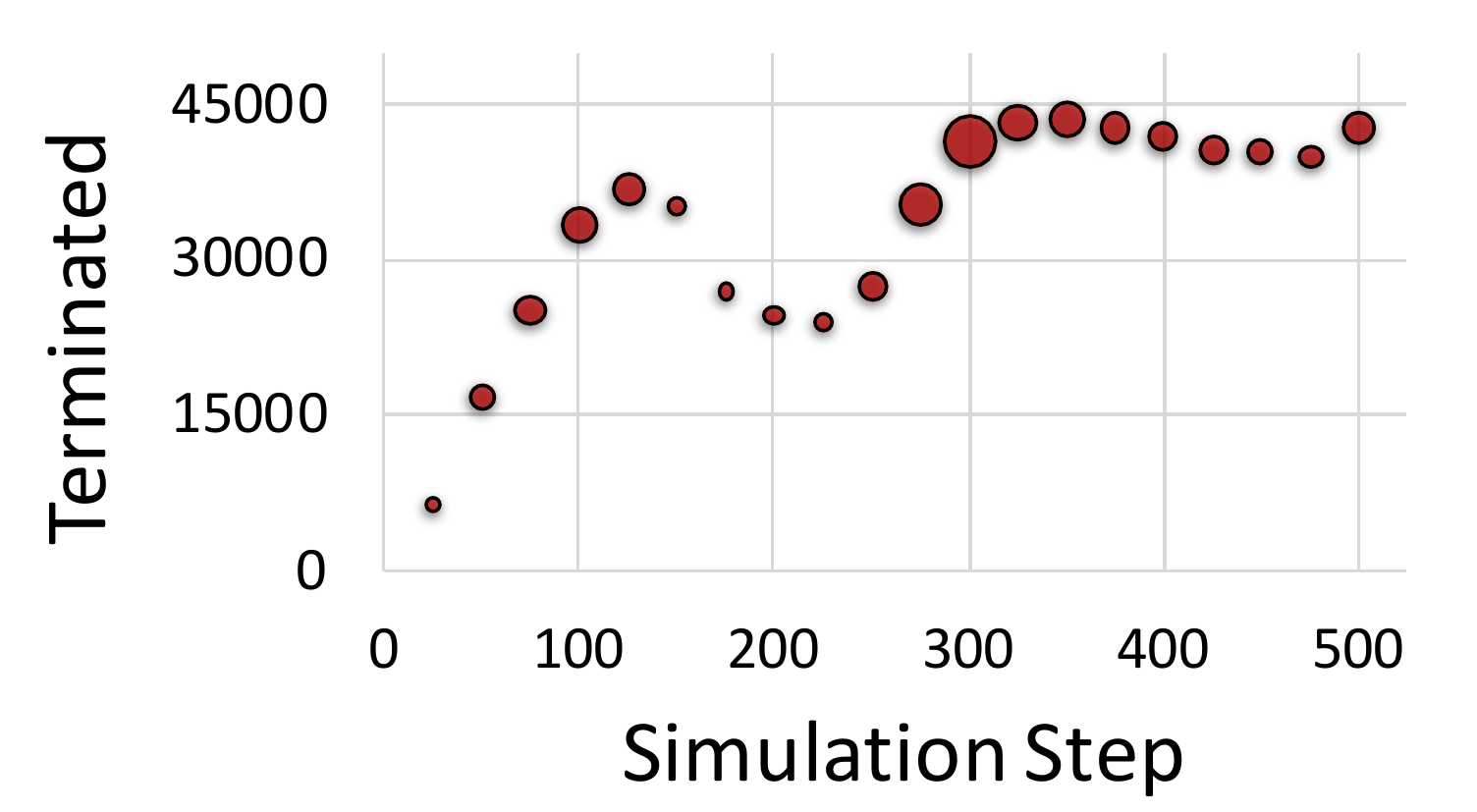}
\caption{Cloverleaf3D}
\label{Clover_Bubble}
\end{subfigure}
\begin{subfigure}{0.245\textwidth}
\raggedright
\includegraphics[width=1.05\linewidth]{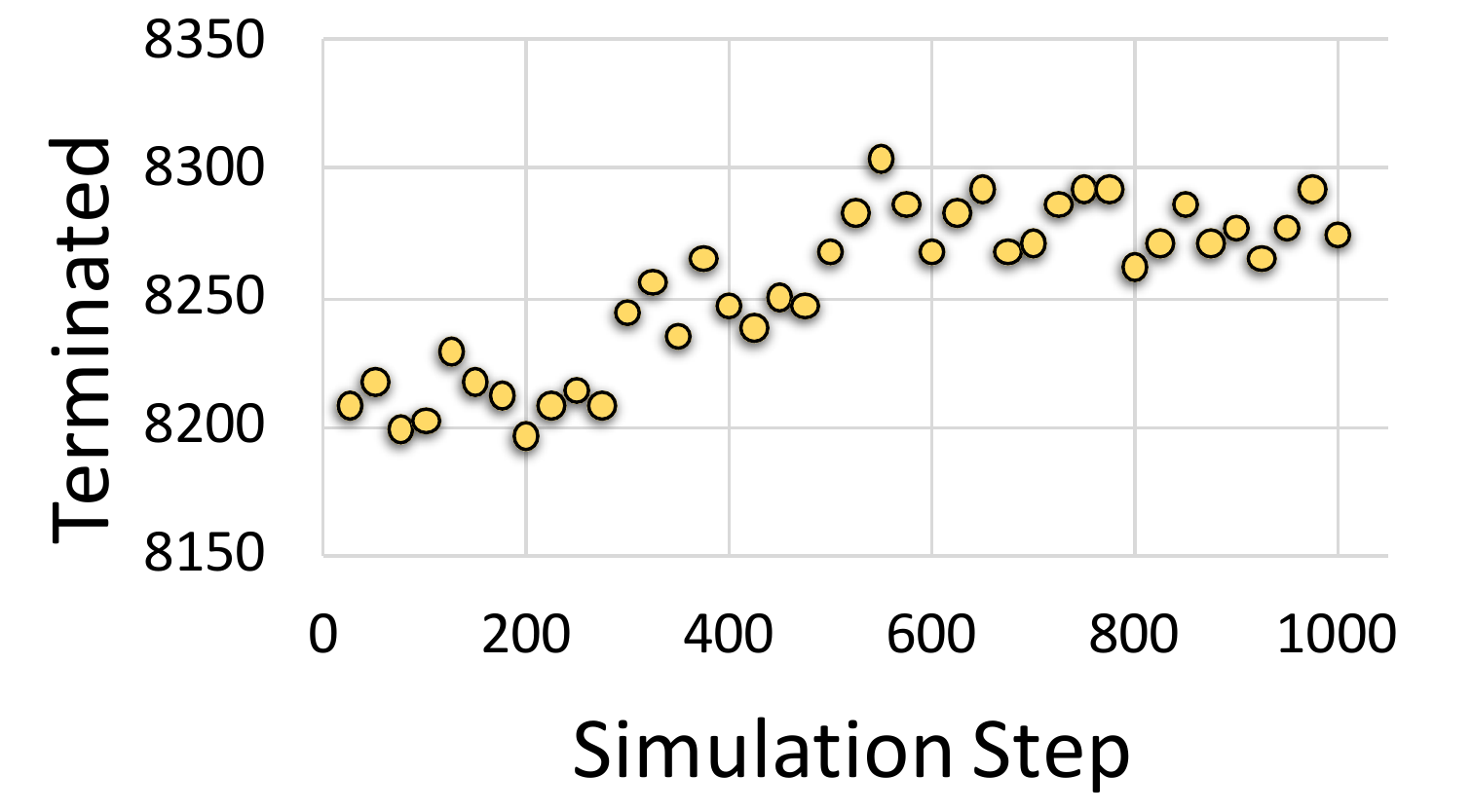}
\caption{ABC}
\label{ABC_Bubble}
\end{subfigure}
\begin{subfigure}{0.245\textwidth}
\raggedright
\includegraphics[width=1.05\linewidth]{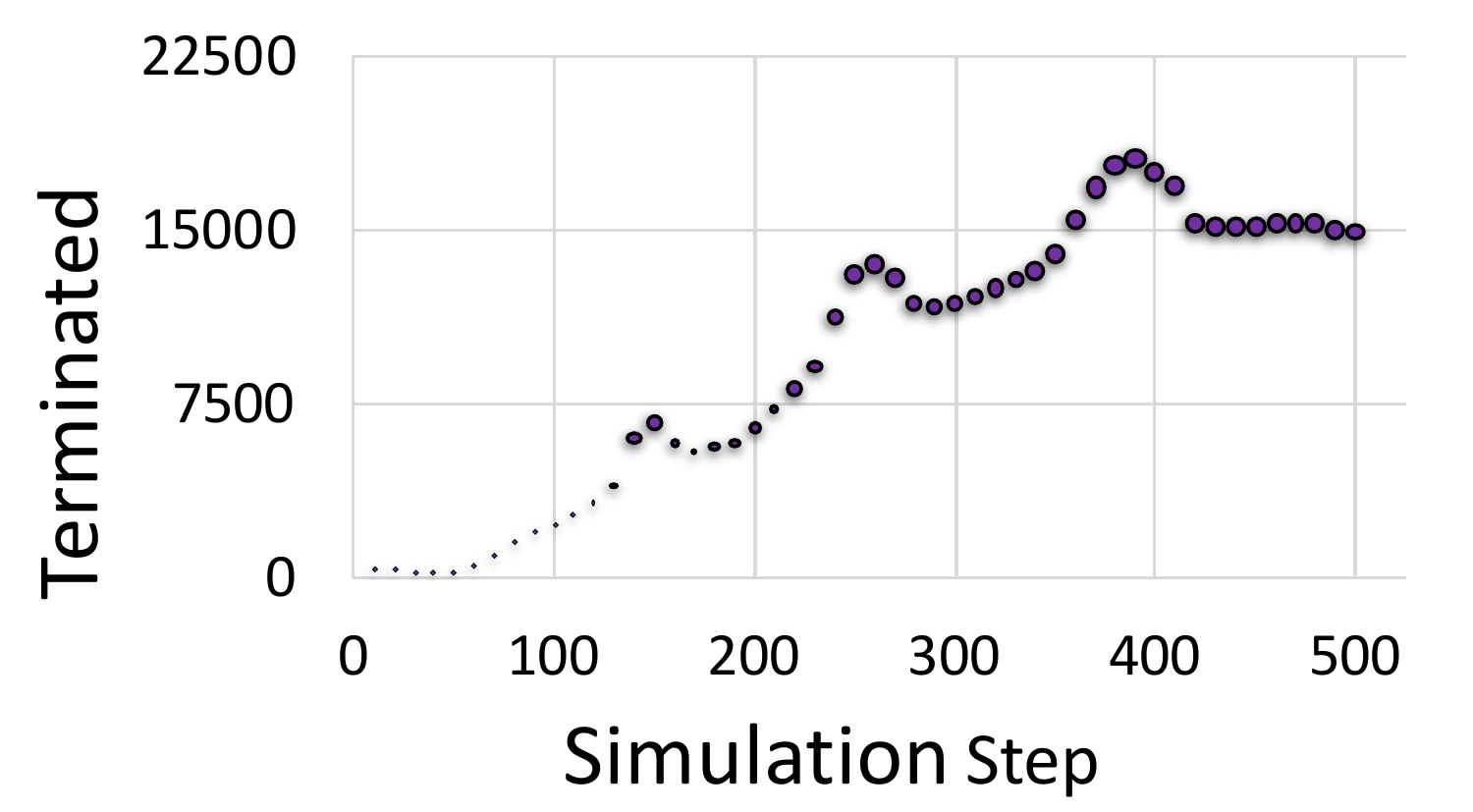}
\caption{Jet Flow}
\label{Jet4_Bubble}
\end{subfigure}
\begin{subfigure}{0.245\textwidth}
\raggedright
\includegraphics[width=1.05\linewidth]{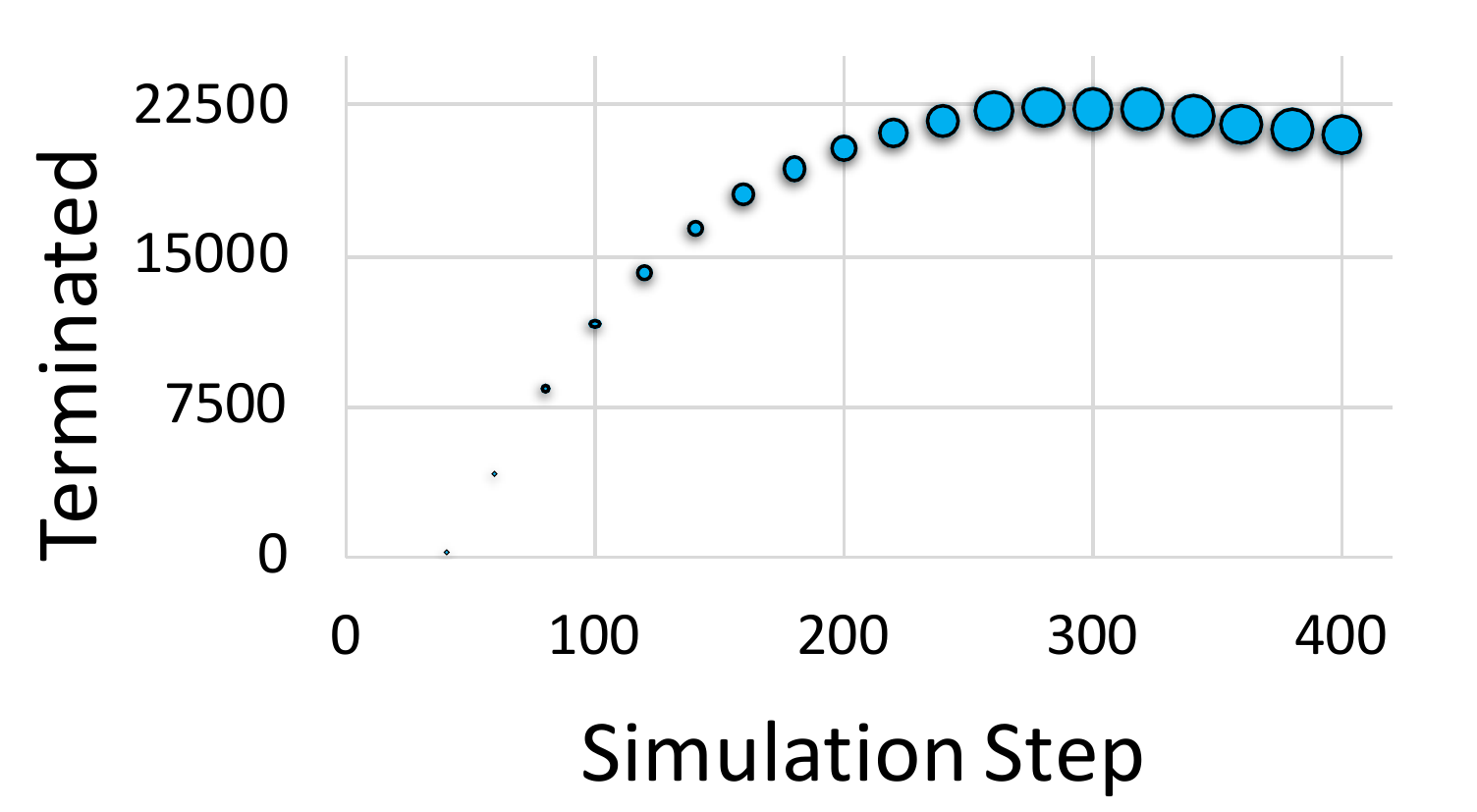}
\caption{Nyx}
\label{Nyx_Bubble}
\end{subfigure}
\begin{subfigure}{0.245\textwidth}
\raggedright
\includegraphics[width=1.05\linewidth]{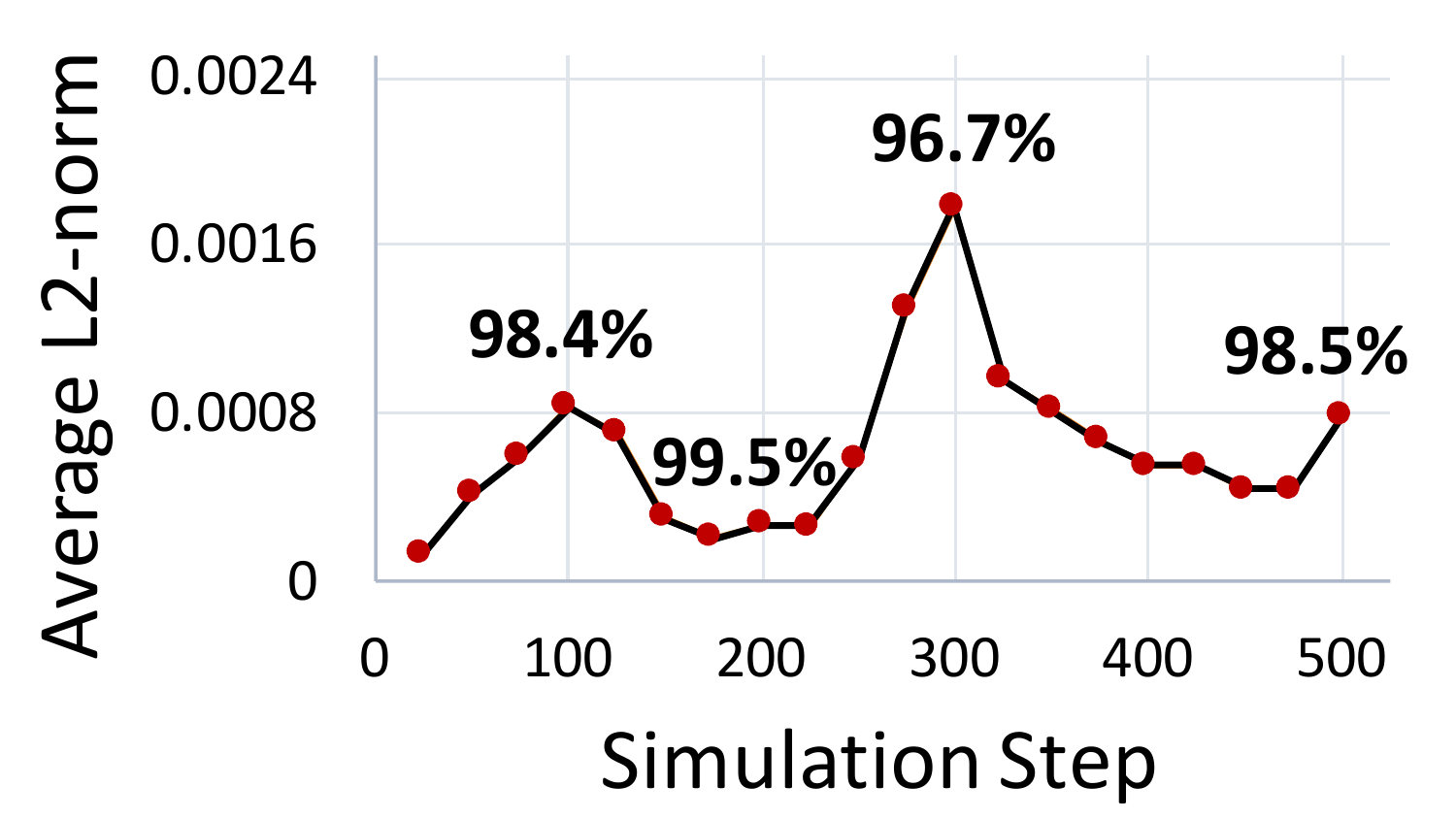}
\caption{Cloverleaf3D}
\label{Clover_Line}
\end{subfigure}
\begin{subfigure}{0.245\textwidth}
\raggedright
\includegraphics[width=1.05\linewidth]{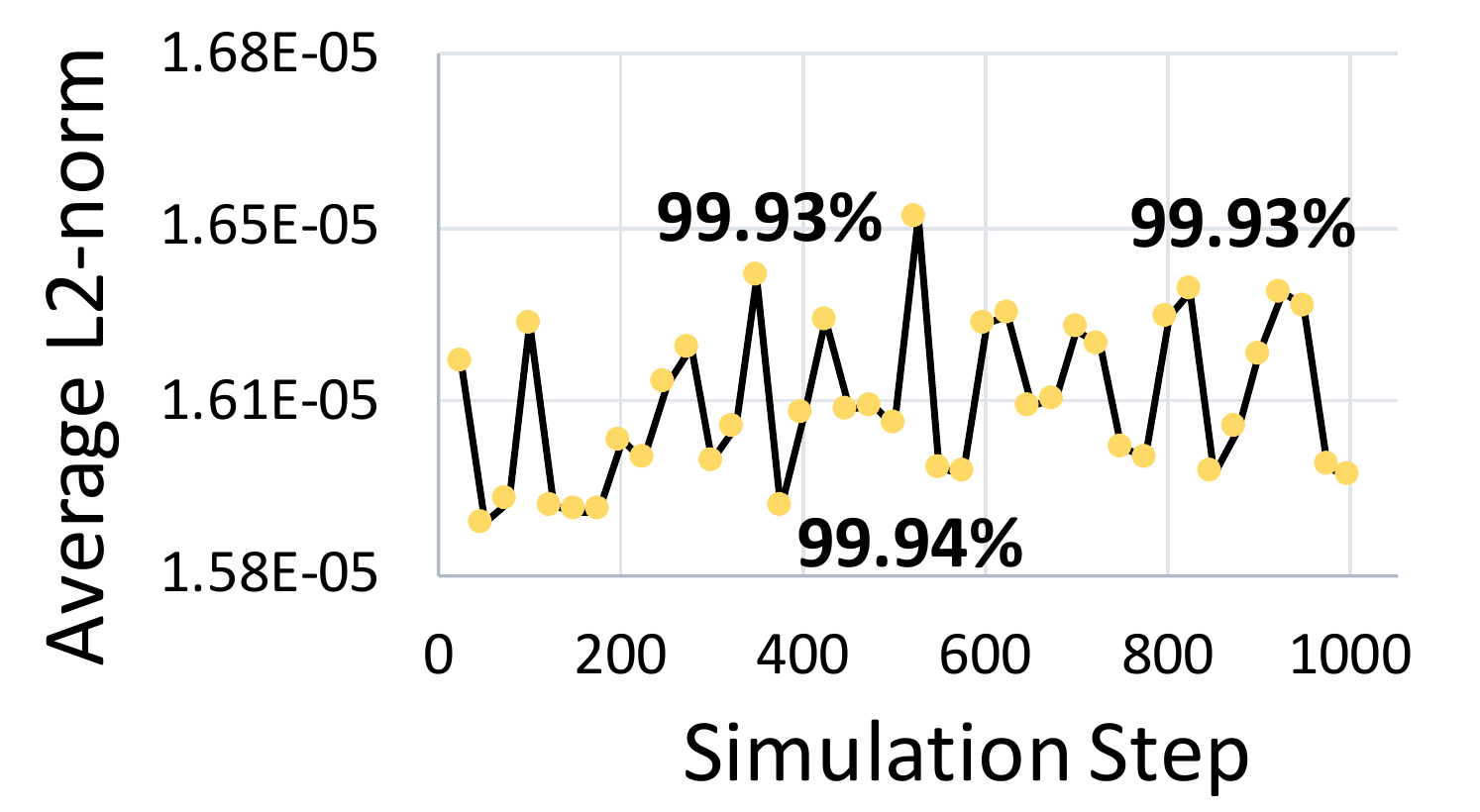}
\caption{ABC}
\label{ABC_Line}
\end{subfigure}
\begin{subfigure}{0.245\textwidth}
\raggedright
\includegraphics[width=1.05\linewidth]{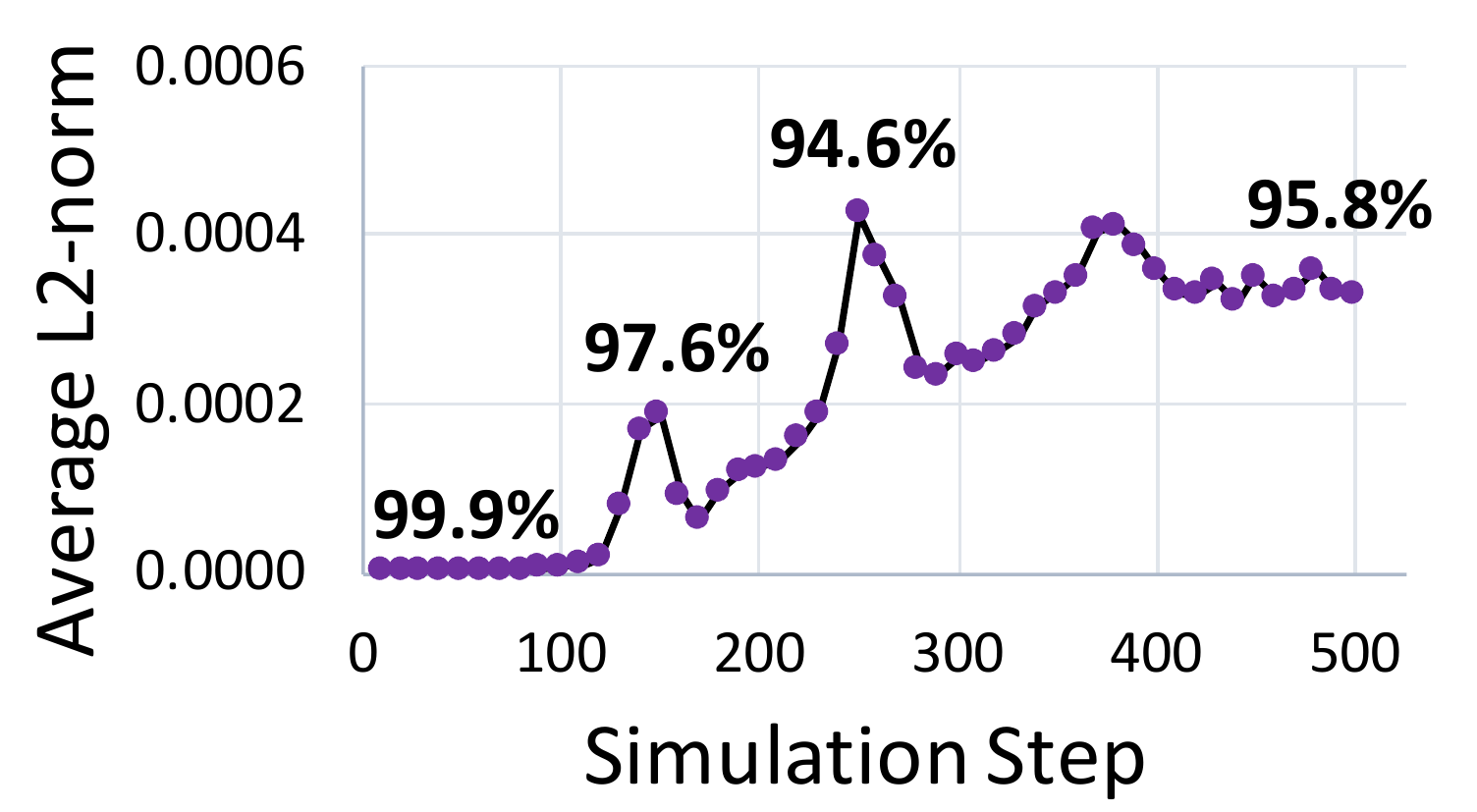}
\caption{Jet flow}
\label{Jet4_Line}
\end{subfigure}
\begin{subfigure}{0.245\textwidth}
\raggedright
\includegraphics[width=1.05\linewidth]{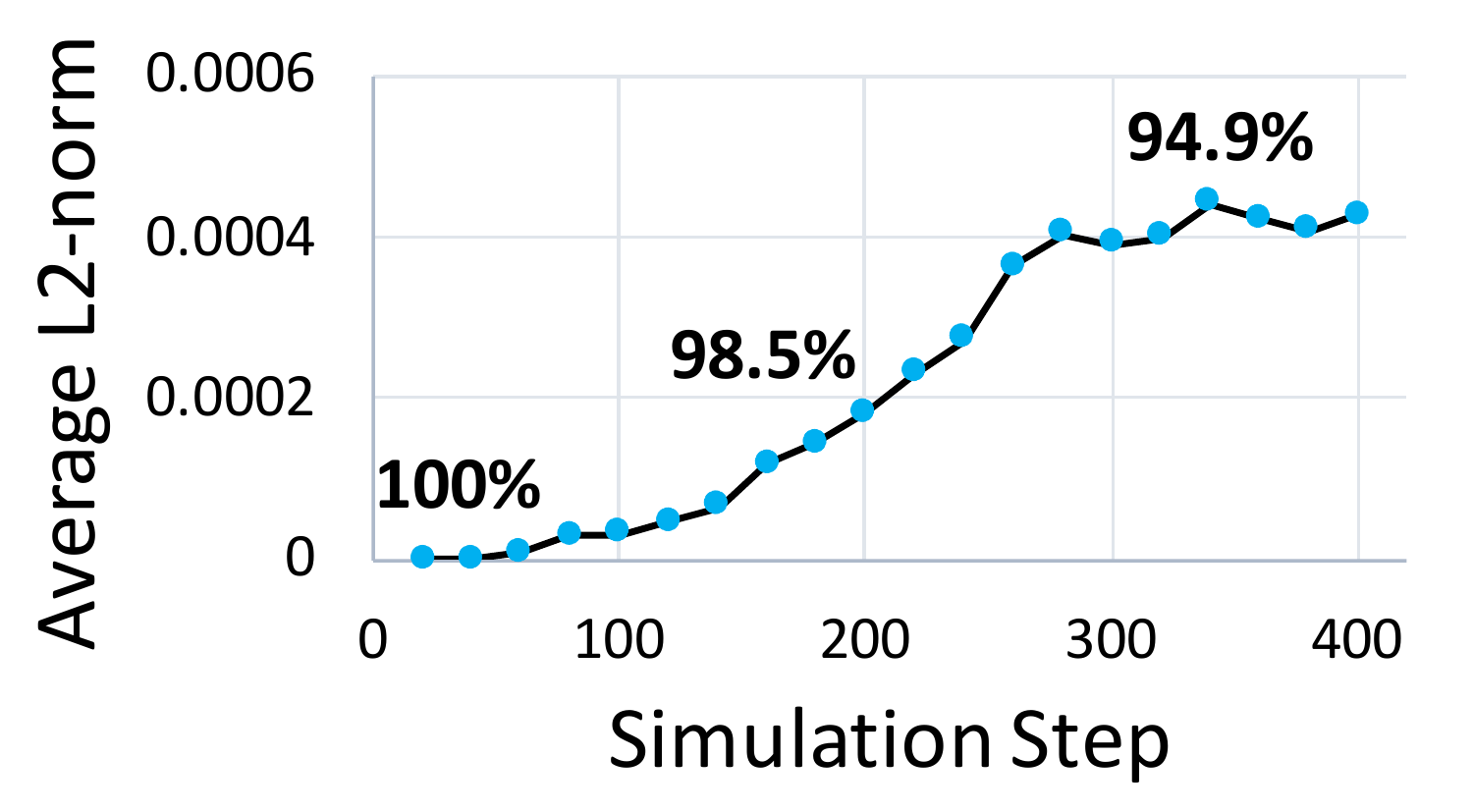}
\caption{Nyx}
\label{Nyx_Line}
\end{subfigure}
\caption{The plots show the relation between number of particles terminated and average L2-norm for all intervals of a single configuration of each data set. 
~\ref{Clover_Bubble} and~\ref{Clover_Line} show row 10 of Table~\ref{cloverleaf_table} for the Cloverleaf3D data set. 
~\ref{ABC_Bubble} and~\ref{ABC_Line} show row 7 of Table~\ref{abc_table} for the ABC data set. 
~\ref{Jet4_Bubble} and~\ref{Jet4_Line} show row 4 of Table~\ref{jet4_table} for the Jet flow data set. 
~\ref{Nyx_Bubble} and~\ref{Nyx_Line} show row 6 of Table~\ref{nyx_table} for the Nyx data set. 
}
\label{dataset_plots}
\end{figure*}

For all data sets, we measured the accuracy of basis flows generated for every interval across all nodes.
Tables~\ref{cloverleaf_table},~\ref{abc_table},~\ref{jet4_table}, and~\ref{nyx_table} show reconstruction accuracy averaged over all the intervals, i.e., average reconstruction accuracy for the entire simulation duration.
Figure~\ref{dataset_plots} shows the change in reconstruction accuracy over every interval and the correlation to the number of particles terminated during that interval for a single configuration of each data set.
Figures~\ref{Clover_Bubble},~\ref{ABC_Bubble},~\ref{Jet4_Bubble}, and~\ref{Nyx_Bubble} use bubble size to represent the number of particles terminated and the corresponding figures~\ref{Clover_Line},~\ref{ABC_Line},~\ref{Jet4_Line}, and~\ref{Nyx_Line} show average L2-norm curves and reconstruction accuracy as a percentage.

\begin{figure*}[!ht]
\begin{subfigure}{0.3\textwidth}
\raggedright
\includegraphics[width=\linewidth,keepaspectratio]{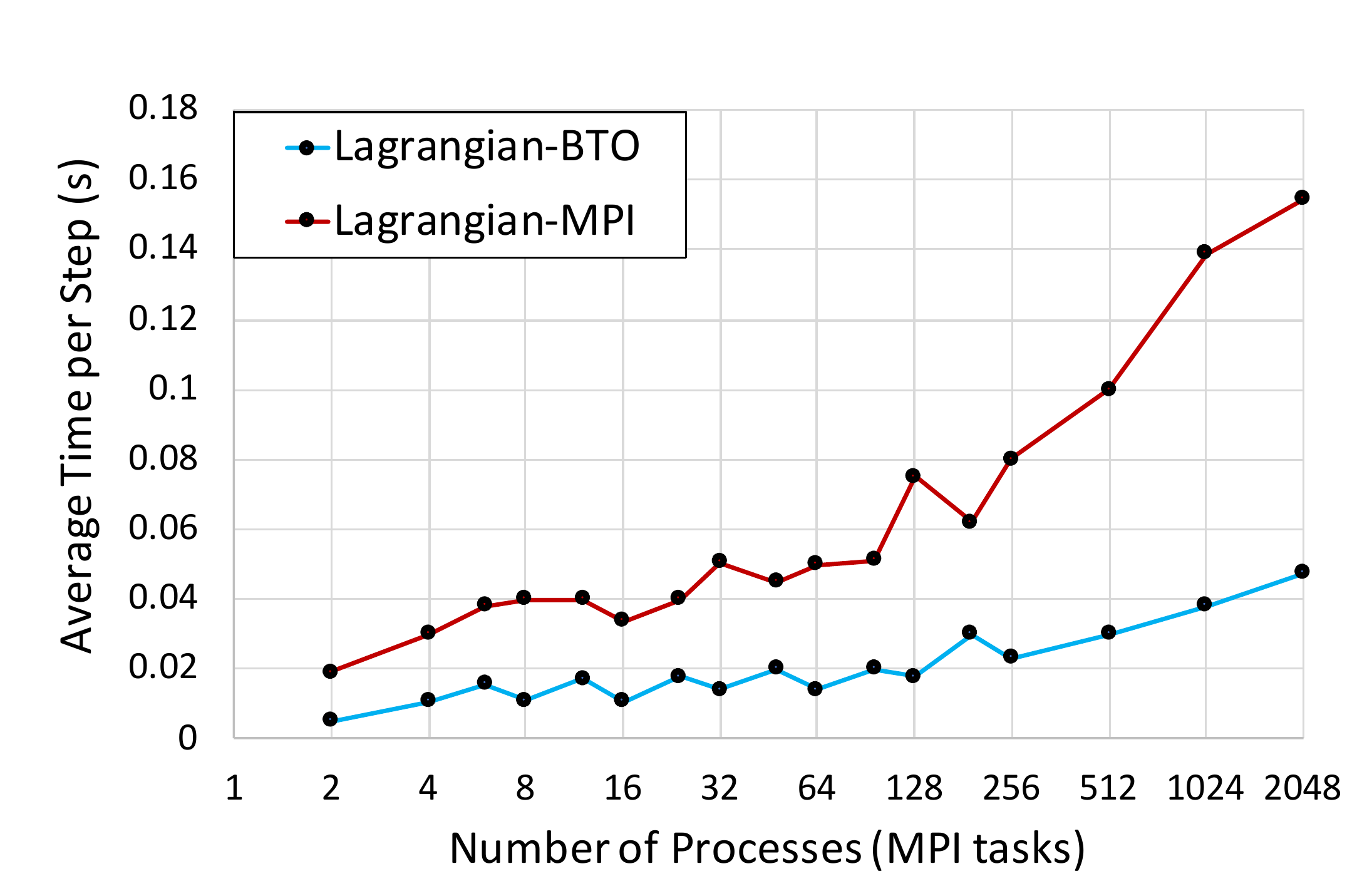}
\caption{}
\label{fig:cloverleaf_weak}
\end{subfigure}
\begin{subfigure}{0.3\textwidth}
\raggedright
\includegraphics[width=\linewidth,keepaspectratio]{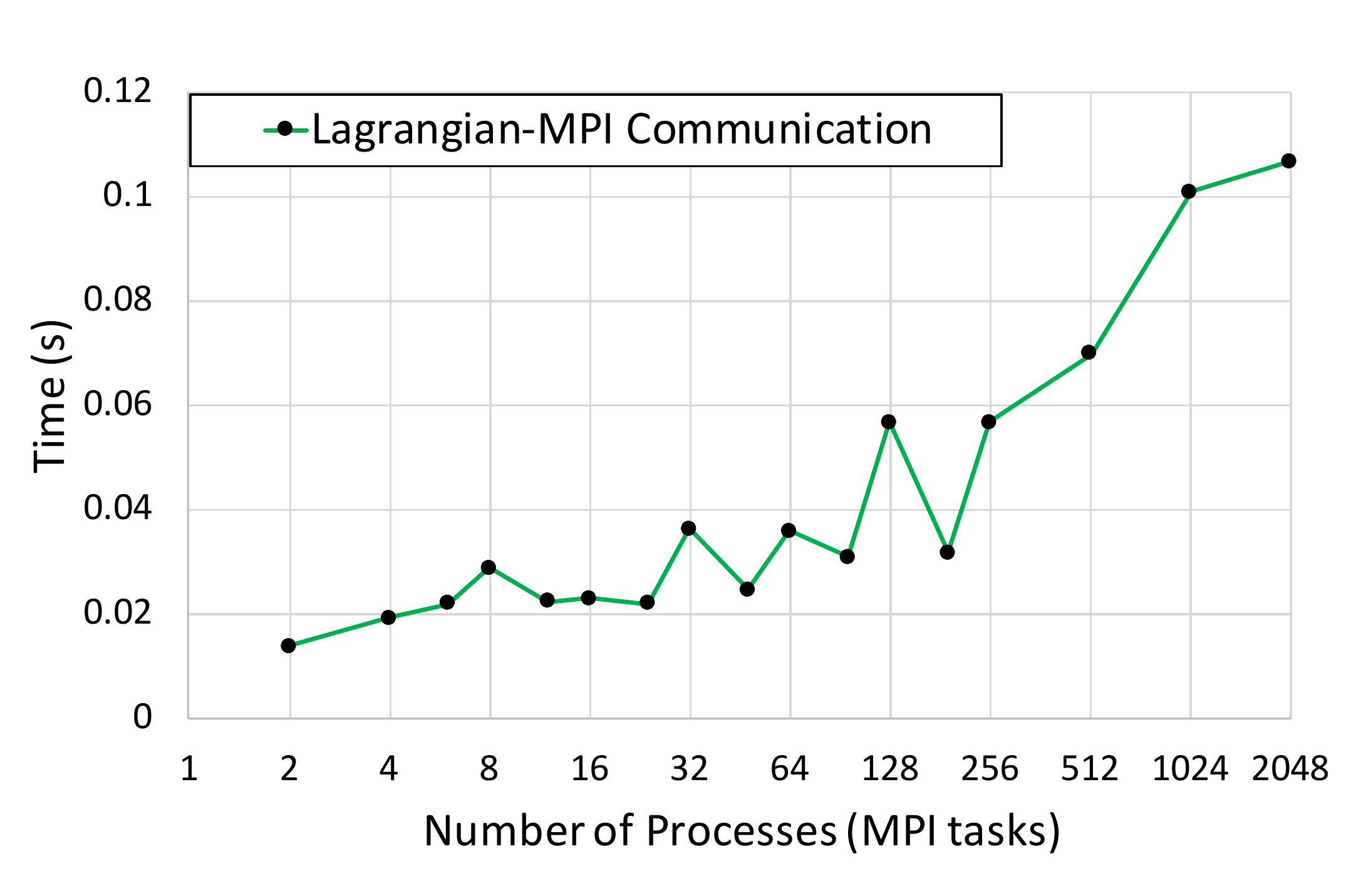}
\caption{}
\label{fig:cloverleaf_time}
\end{subfigure}
\begin{subfigure}{0.17\textwidth}
\centering
\includegraphics[width=1\linewidth,keepaspectratio]{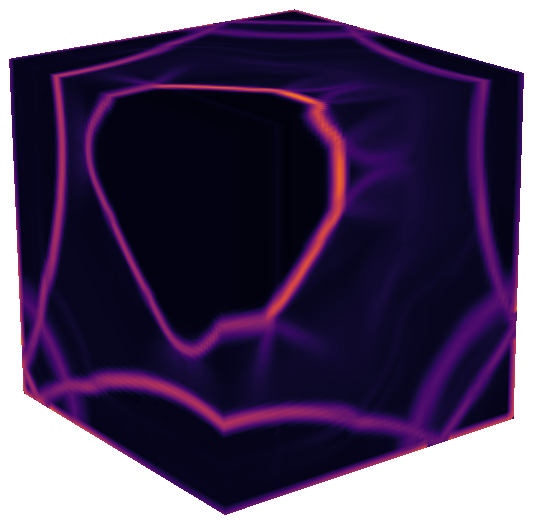}
\caption{Lagrangian-MPI}
\label{clover_mpi_backward}
\end{subfigure}
\begin{subfigure}{0.21\textwidth}
\centering
\includegraphics[width=1\linewidth,keepaspectratio]{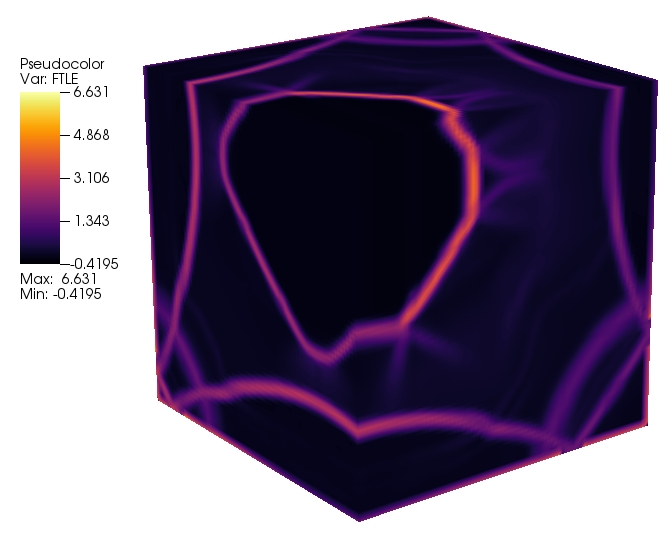}
\caption{Lagrangian-BTO}
\label{clover_bto_backward}
\end{subfigure}
\caption{Weak scaling study results for the Cloverleaf3D data set. In~\ref{fig:cloverleaf_weak} each curve plots the average time required for a single step for an increasing number of MPI tasks. Lagrangian-BTO is on average 3x faster than Lagrangian-MPI. \ref{fig:cloverleaf_time} shows the approximate time required for MPI communication by the Lagrangian-MPI analysis filter averaged across all processes. All measurements are in seconds. ~\ref{clover_mpi_backward} and~\ref{clover_bto_backward} compare FTLE visualizations generated post hoc using basis flows from Lagrangian-MPI and Lagrangian-BTO, respectively.}
\label{fig:clover_ftle_visualizations}
\end{figure*}

To supplement our quantitative evaluation, Figures~\ref{clover_mpi_backward},~\ref{clover_bto_backward},~\ref{fig:abc_ftle_visualizations},~\ref{fig:jet_ftle_visualizations}, and~\ref{fig:nyx_ftle_visualizations} provide a qualitative comparison using colormapped visualizations of surfaces of subvolumes of FTLE scalar fields generated post hoc using basis flows calculated by Lagrangian-MPI and Lagrangian-BTO.
Well-defined ridges in the FTLE field~(identified by high scalar values), used to visualize Lagrangian Coherent Structures, are of particular interest in these figures. 

\begin{table*}[!h]
\centering
\scalebox{0.75}{
\begin{tabular}{||c|c|c|c|c||c|c|b||c|c|c|c|d||}
\hline
\textbf{Nodes} & \textbf{MPI} & \textbf{GPUs/} & \textbf{Dims} & \textbf{Step} & \textbf{L-BTO~(s)} & \textbf{L-MPI~(s)} & \textbf{Speed-up} & \textbf{Discarded \%} & \textbf{Greatest Max} & \textbf{Average Max} & \textbf{Total Average} & \textbf{Accuracy \%}\\
& \textbf{Ranks} & \textbf{Node} &  & \textbf{Size}  &  &  &  &  & \textbf{L2-norm} & \textbf{L2-norm} & \textbf{L2-norm} & \\
\hline
1 & 2 & 2 & $81^3$ & 0.038 & 0.0050 & 0.0190 & \textbf{3.8x} & 1.9 & 2.52$\times10^{-3}$ & 7.58$\times10^{-4}$  & 4.53$\times10^{-4}$ & \textbf{99.6} \\
1 & 4 & 4 & $102^3$ & 0.029 & 0.0106 & 0.0301 & \textbf{2.8x} & 2  & 2.44$\times10^{-3}$ & 7.68$\times10^{-4}$  & 4.19$\times10^{-4}$ & \textbf{99.5} \\
1 & 6 & 6 & $116^3$ & 0.025 & 0.0158 & 0.0380 & \textbf{2.4x} & 2.8  & 3.43$\times10^{-3}$ &  1.20$\times10^{-3}$ & 4.89$\times10^{-4}$ & \textbf{99.4} \\
\hline
2 & 8 & 4 & $128^3$ & 0.023 & 0.0109 & 0.0405 & \textbf{3.7x} & 1.8  & 1.61$\times10^{-3}$ &  6.49$\times10^{-4}$  & 2.60$\times10^{-4}$ & \textbf{99.6} \\
2 & 12 & 6 & $146^3$ & 0.019 & 0.0173 & 0.0398 & \textbf{2.3x} & 2.4  & 2.84$\times10^{-3}$ & 1$\times10^{-3}$  & 3.09$\times10^{-4}$ & \textbf{99.5} \\
\hline
4 & 16 & 4 & $161^3$ & 0.017 & 0.0107 & 0.0338 & \textbf{3.1x} & 2.9  & 1.11$\times10^{-2}$ &  1.97$\times10^{-3}$ & 4.38$\times10^{-4}$ & \textbf{99.2} \\
4 & 24 & 6 & $184^3$ & 0.015 & 0.0178 & 0.0410 & \textbf{2.3x} & 4.6  & 5.83$\times10^{-3}$ & 2.35$\times10^{-3}$  & 6.28$\times10^{-4}$ & \textbf{98.8} \\
\hline
8 & 32 & 4 & $203^3$ & 0.013 & 0.0140 & 0.0506 & \textbf{3.6x} & 4.1  & 3.11$\times10^{-3}$ & 1.44$\times10^{-3}$  & 3.83$\times10^{-4}$ & \textbf{99.2} \\
8 & 48 & 6 & $232^3$ & 0.011 & 0.0201 & 0.0449 & \textbf{2.2x} & 4.9  & 5.70$\times10^{-3}$ & 3.15$\times10^{-3}$  & 4.45$\times10^{-4}$ & \textbf{98.9} \\
\hline
16 & 64 & 4 & $256^3$ & 0.010 & 0.0140 & 0.0504 & \textbf{3.6x} & 4.5  & 8.39$\times10^{-3}$ & 3.58$\times10^{-3}$  & 3.62$\times10^{-4}$ & \textbf{99.1} \\
16 & 96 & 6 & $293^3$ & 0.009 & 0.0200 & 0.0510 & \textbf{2.5x} & --- & --- & --- & --- & --- \\
\hline
32 & 128 & 4 & $322^3$ & 0.008 & 0.0180 & 0.0750 & \textbf{4.1x} & --- & --- & --- & --- & --- \\
32 & 192 & 6 & $370^3$ & 0.007 & 0.0301 & 0.0620 & \textbf{2.0x} & --- & --- & --- & --- & --- \\
\hline
64 & 256 & 4 & $406^3$ & 0.006 & 0.0230 & 0.0808 & \textbf{3.5x} & --- & --- & --- & --- & --- \\
\hline
128 & 512 & 4 & $512^3$ & 0.005 & 0.0303 & 0.1001 & \textbf{3.3x} & --- & --- & --- & --- & --- \\
\hline
256 & 1024 & 4 & $645^3$ & 0.004 & 0.0380 & 0.1390 & \textbf{3.6x} & --- & --- & --- & --- & --- \\
\hline
512 & 2048 & 4 & $812^3$ & 0.003 & 0.0475 & 0.1544 & \textbf{3.2x} & --- & --- & --- & --- & --- \\
\hline
\end{tabular}}
\caption{Weak scaling configurations and timing results for the Cloverleaf3D data set \textbf{WF1} experiments. Lagrangian-BTO and Lagrangian-MPI columns show the average time per step in seconds. We post-processed and measured accuracy for only a subset of the WF1 experiments (10 of 17), since calculating the reconstruction accuracy takes prohibitively long periods of time on our local cluster. Each node operates on an approximately $64^{3}$ grid. Reported results are measured across all intervals.}
\label{cloverleaf_table}
\end{table*}

\subsection{\textbf{Cloverleaf3D Simulation}}
\label{sec:clover}
Cloverleaf3D is a three-dimensional version of the Lagrangian-Eulerian explicit hydrodynamics mini-application Cloverleaf~\cite{mallinson2013cloverleaf}. 
It has been developed and used to evaluate techniques targeting Exascale applications.
%
%
%

\subsubsection{Weak Scaling}
We performed \textbf{WF1} experiments, i.e., an in situ weak scaling study to evaluate performance, using the Cloverleaf3D simulation on Summit. 
For each run, we terminated the simulation after 500 cycles.
Given that our study varies the resolution of the simulation with the number of processes, the total number of simulation steps varies. 
4 of 17 experiment configurations completed before 500 cycles and all the simulations that are terminated at 500 cycles reach different stages of the simulation~(variable step size).
In general, the greater the spatial resolution of the data set, the greater the number of cycles required to reach completion.
Our experiment configurations span $81^3$ across 2 MPI tasks to $812^3$ across 2048 MPI tasks, with each MPI rank operating on an approximately $64^3$ grid.
In each case, a single MPI task is allocated a single GPU.
Thus, our smallest configuration used 2 GPUs on a single compute node, and the largest used 2048 GPUs across 512 compute nodes.
Additionally, given that each node on Summit has 6 GPUs, we varied the number of GPUs utilized on a single node to gauge on-node MPI communication optimizations and performance of particle advection using multiple GPUs on the same node.
For each experiment, we set the maximum step size to 0.1. 
However, Cloverleaf3D uses an adaptive step size based on the simulation grid resolution.
We report the average step size taken by the simulation, configuration information~(number of nodes, MPI ranks, and GPUs per node, and dimensions), and scalability performance results in Table~\ref{cloverleaf_table}. 
Further, each weak scaling test extracted basis flows at an interval of every 25 cycles and used a data reduction of 1:8, i.e., one particle for every eight grid points.

\textbf{Performance ---} Figure~\ref{fig:cloverleaf_weak} compares the average time required per step by each technique.
Lagrangian-BTO scales better than Lagrangian-MPI because it is a communication-free analysis filter.
%
For the extraction of Lagrangian flow maps in situ, Lagrangian-BTO demonstrates an average of 3x speed-up over Lagrangian-MPI. 
We observe the cost of particle advection (for both techniques) increases as the scale of the simulation increases.
However, each process operates on an approximately $64^3$ grid irrespective of the total number of MPI tasks, and each Lagrangian-BTO process operates without any knowledge of other processes.
In addition to the number of particles being advected, multiple variables influence the cost of the particle advection worklet, namely, cell size, step size, and the vector field.
Performing Runge Kutta fourth-order interpolation could require interpolating velocity information from multiple cells (determined by cell size, step size, and the vector field).
Although the increased cost of particle advection affects the time required by both Lagrangian analysis techniques, 
%
exact comparisons between different configurations (i.e., rows in Table~\ref{cloverleaf_table}) would not account for the above parameters.
We believe the performance of the particle advection worklet considering the above parameters is worth future investigation.
Further, use of a faster particle advection kernel would result in greater speed-ups for the Lagrangian-BTO technique.

\textbf{Varying number of GPUs/node ---}
The ``sawtooth'' nature of the plots in Figures~\ref{fig:cloverleaf_weak} and \ref{fig:cloverleaf_time} is a result of alternating the number of GPUs being utilized between 4 and 6. 
Particle advection performs better with 4 GPUs per node versus 6. 
The use of shared memory by multiple GPUs on a single node and saturation of the NVLink by the VTK-m particle advection worklet causes this effect.
Figure~\ref{fig:cloverleaf_time} captures the difference in time between both curves, i.e., approximates the time spent on communication, and shows a reduction in MPI communication costs when using 6 ranks versus 4 per node, albeit scaling poorly as the number of nodes increases.
On-node MPI communication optimizations contribute to better performance when grouping a larger number of MPI tasks on each node. 
However, 
as the number of nodes increases, the cost of inter-node communication remains high in comparison to on-node communication.

\textbf{Accuracy ---} Higher resolution simulation configurations use a proportionately larger number of particles and thus generate more basis flows.
We measured accuracy for only a subset of the WF1 experiments (10 of 17), since calculating the reconstruction accuracy takes prohibitively long periods of time on our local cluster. 
Lagrangian-BTO terminated between 2-5\% of basis flows on average across all experiments, i.e., 2-5\% less data was saved to disk.
However, after we interpolate the missing trajectories using the saved basis flows, we approximate the complete flow map (as generated by Lagrangian-MPI) with over 99\% accuracy on average using the L2-norm metric.
%
%

%

%

Figures~\ref{clover_mpi_backward} and~\ref{clover_bto_backward} compare FTLE visualizations generated using approximately 2M basis flows saved by each technique (configuration in row 10 of Table~\ref{cloverleaf_table}). 
We observe no significant differences in the visualizations with both methods capturing the FTLE ridges with the same quality.

%

Overall, our weak scaling study showed that for the Cloverleaf3D data set, Lagrangian-BTO was on average 3x faster, while generating 99\% as accurate flow maps on average and qualitatively comparable post hoc FTLE visualizations as Lagrangian-MPI. 
%
%
In situ analysis using Lagrangian-MPI contributed between 5-12\% of the total simulation time, and in most cases, Lagrangian-BTO required 50\%-80\% less time than the corresponding Lagrangian-MPI configuration.

\begin{figure}[!b]
\raggedright
\includegraphics[width=1.03\linewidth]{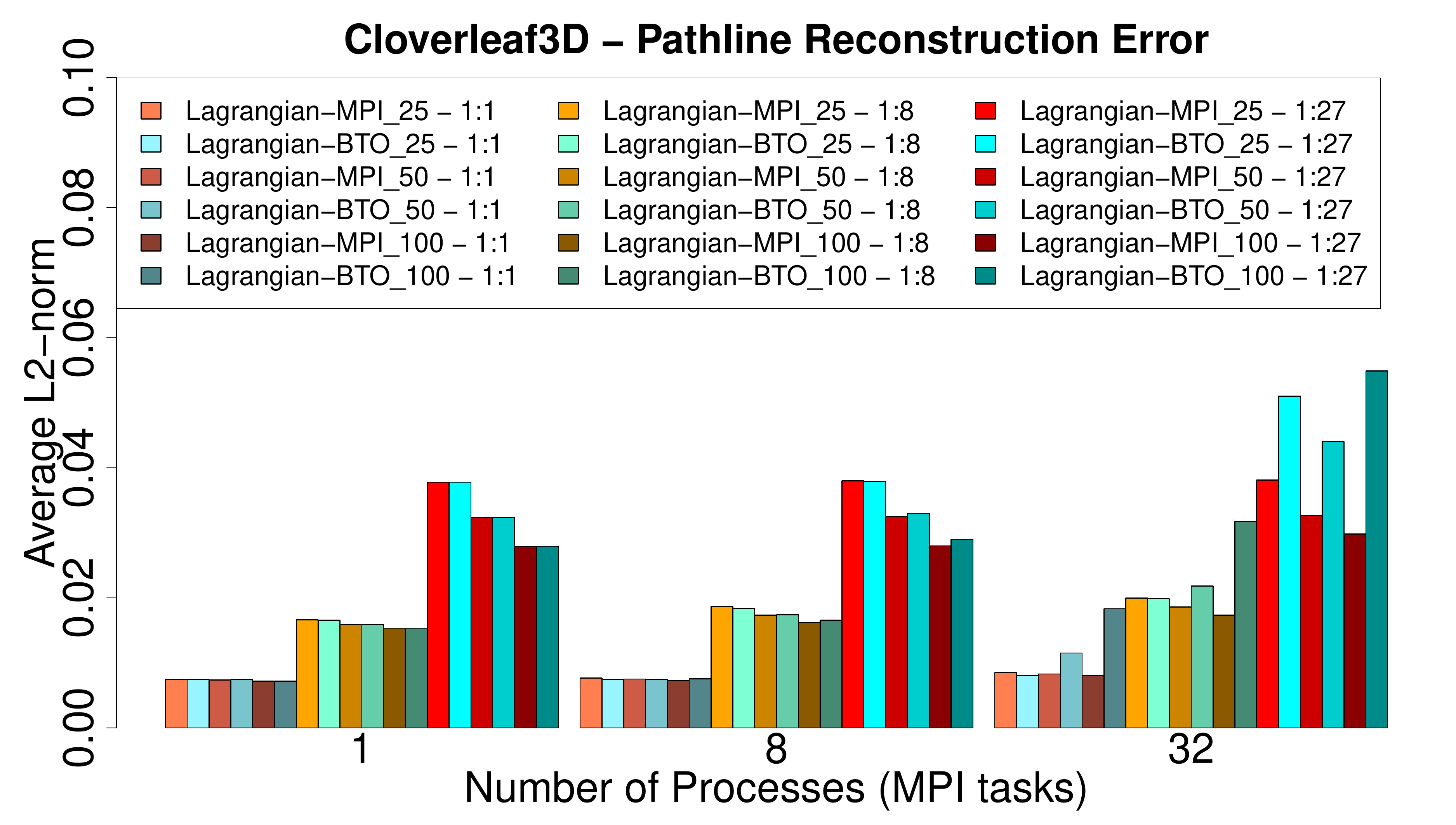}
\caption{Strong scaling pathline reconstruction error plot for the Cloverleaf3D data set \textbf{WF2} experiments. 
We use warm colored bars for Lagrangian-MPI and cool colored bars for Lagrangian-BTO. 
Bars are grouped from left to right by number of MPI tasks used, i.e., in increasing order of degree of domain decomposition.
Within each group of bars, configurations are subgrouped by data reduction options, i.e, 1:1, 1:8, and 1:27.
Within a subgroup, pairs of matching Lagrangian-MPI and Lagrangian-BTO tests are ordered from left to right by interval value.
In these experiments, a $64^3$ dimension data set was used. The total number of simulation cycles was set to 800. 
We traced pathlines for 800 cycles using the basis flows generated by each Lagrangian analysis technique and compared them to the ground truth.
}
\label{fig:clover_strong}
\end{figure}

\subsubsection{Strong Scaling} 
Whereas the weak scaling study demonstrated the viability of using Lagrangian-BTO, the strong scaling study helps us understand its limitations.
We considered a $64^3$ grid and decomposed it into smaller grids based on the number of MPI tasks.
Further, for these \textbf{WF2} experiments, we considered 
interval values of 25, 50, and 100, and data reduction values of 1:1, 1:8, and 1:27. 
The interval determines how long particles advect before their end locations are saved.
The longer the interval, the greater chance of particles exiting the node domain and vice versa.
Further, the number of particles that exit the node domain and terminate directly impacts the accuracy of the approximation.
Figure~\ref{fig:clover_strong} shows the accuracy results when decomposing the domain across 1, 8, and 32 MPI tasks. 
%

\begin{table*}[!ht]
\centering
\scalebox{0.8}{
\begin{tabular}{||c|c||c|c|b||c|c|c|c|d||}
\hline
\textbf{Interval} & \textbf{Reduction} & \textbf{L-BTO~(s)} & \textbf{L-MPI~(s)} & \textbf{Speed-up} & \textbf{Discarded \%} & \textbf{Greatest Max} & \textbf{Average Max} & \textbf{Total Average} & \textbf{Accuracy \%} \\
 &  &  &  & & & \textbf{L2-norm} & \textbf{L2-norm} & \textbf{L2-norm} & \\
\hline
25 & 1:1 & 0.0059 & 0.0143 & \textbf{2.3x} & 2.7 & 6.85$\times10^{-4}$ & 4.97$\times10^{-4}$  & 6.15$\times10^{-5}$ & \textbf{99.7} \\
\hline
50 & 1:1 & 0.0066 & 0.0145 & \textbf{2.1x} & 5.5 & 3.03$\times10^{-3}$ & 1.85$\times10^{-3}$  & 2.31$\times10^{-4}$ & \textbf{99.0} \\
\hline
100 & 1:1 & 0.0067 & 0.0144 & \textbf{2.1x} & 10.9 & 7.87$\times10^{-3}$ & 6.81$\times10^{-3}$  & 1.09$\times10^{-3}$ & \textbf{95.5} \\
\hline
25 & 1:8 & 0.0026 & 0.0065 & \textbf{2.4x} & 2.6 & 1.40$\times10^{-3}$ & 3.41$\times10^{-4}$  & 3.93$\times10^{-5}$ & \textbf{99.8} \\
\hline
50 & 1:8 & 0.0033 & 0.0086 & \textbf{2.6x} & 5.4 & 3.71$\times10^{-3}$ & 1.23$\times10^{-3}$ & 1.32$\times10^{-4}$ & \textbf{99.4} \\
\hline
100 & 1:8 & 0.0026 & 0.0075 & \textbf{2.8x} & 10.8 & 6.15$\times10^{-3}$ & 2.63$\times10^{-3}$ & 3.65$\times10^{-4}$ & \textbf{98.5} \\
\hline
25 & 1:27 & 0.0024 & 0.0065 & \textbf{2.7x} & 1.6 & 5.24$\times10^{-5}$ & 4.87$\times10^{-5}$ & 1.61$\times10^{-5}$ & \textbf{99.9} \\
\hline
50 & 1:27 & 0.0029 & 0.0048 & \textbf{1.6x} & 5.1 & 2.62$\times10^{-3}$ &  1.28$\times10^{-3}$ & 1.23$\times10^{-4}$ & \textbf{99.5} \\
\hline
100 & 1:27 & 0.0027 & 0.0053 & \textbf{1.9x} & 10.4 & 1.25$\times10^{-2}$ &  6.53$\times10^{-3}$ & 5.01$\times10^{-4}$ & \textbf{97.9} \\
\hline
\end{tabular}
}
\caption{Speed-up and reconstruction accuracy results for the ABC data set \textbf{WF2} experiments. Lagrangian-BTO and Lagrangian-MPI columns show average time per step in seconds. Reported results are measured across all intervals.}
\label{abc_table}
\end{table*}

For these experiments, we measured accuracy by calculating pathlines using the basis flows generated by each technique and evaluating their accuracy in comparison to a ground truth at interpolated locations.
The ground truth is calculated by performing RK4 advection using the full spatial and temporal resolution of the data set.
We placed 1,000 particles in the domain to generate pathlines for this evaluation.
We compared the similarity between ground truth and pathlines generated by interpolating Lagrangian basis flows using the average L2-norm metric.
Our metric compared the accuracy of interpolated points along the trajectory~\cite{sane2019interpolation}.
Lastly, for these experiments, we loaded Cloverleaf3D vector field data from disk and set the maximum simulation step size to 0.01 and used 800 simulation steps. 
Therefore, we generated pathlines for a duration of 800 simulation steps by stitching together results using successive batches of basis flows~\cite{agranovsky2014improved,bujack2015lagrangian,sane2018revisiting}.

Figure~\ref{fig:clover_strong} groups configurations by number of processes used, i.e., the degree of decomposition.
We note that Lagrangian-MPI accuracy is not affected by degree of decomposition and hence remains constant irrespective of number of processors.
When considering only a single processor, both methods lose particles that exit the entire domain during the interval of advection.
Thus, performance on a single MPI task is identical for both methods.
When decomposed across 8 tasks, we see the error of Lagrangian-BTO increases slightly for configurations that use a 1:27 data reduction.
Specifically, accuracy reduces from 100\% as accurate as Lagrangian-MPI to 96\% as accurate when the interval increases from 25 to 100. 
For domain decomposition across 32 tasks, i.e., the highest degree of decomposition we consider, Lagrangian-BTO error increases as the interval increases and fewer particles are used.
%
%
These experiments are valuable in demonstrating the limitations of the Lagrangian-BTO method.
However, we believe our strong scaling study tests an extreme case, because we expect our target applications will have higher resolution grids 
and use a larger number of particles to capture the flow field behavior.
%


\subsection{\textbf{Arnold-Beltrami-Childress (ABC)}}
\label{sec:abc}
We used a 3D time-dependent variant of the ABC flow analytic vector field~\cite{brummell2001linear}. 
We used one complete period of the flow for a total of 1000 time steps at a grid resolution of $256^3$.
For the \textbf{WF2} ABC data set experiments, we considered three options for interval~(25, 50, 100) and three options for data reduction~(1:1:, 1:8, 1:27).
All tests use 16 nodes, 64 MPI tasks, with 4 MPI tasks using 4 GPUs on each node.
Table~\ref{abc_table} contains configuration details, percentage of discarded particles, speed-up achieved, and the reconstruction accuracy percentages for the ABC data set.

\begin{figure}[!b]
\hspace{4mm}
\begin{subfigure}{0.36\linewidth}
\centering
\includegraphics[width=1\linewidth]{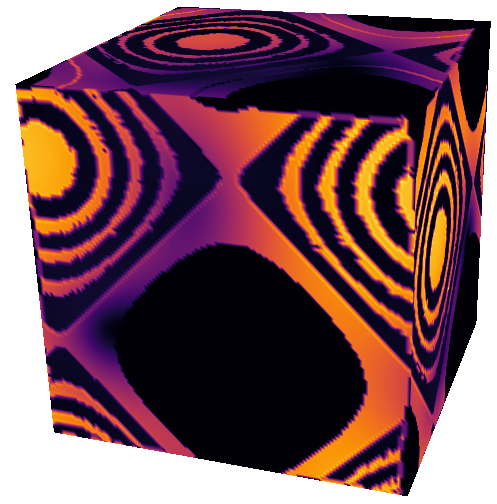}
\caption{Lagrangian-MPI}
\label{abc_mpi_backward}
\end{subfigure}
\hspace{2mm}
\begin{subfigure}{0.44\linewidth}
\centering
\includegraphics[width=1\linewidth]{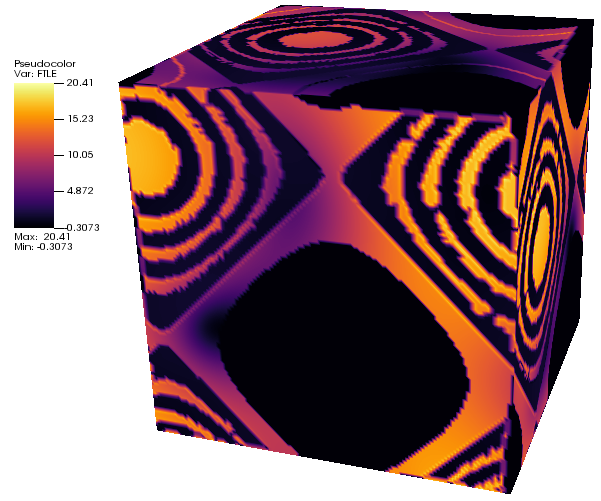}
\caption{Lagrangian-BTO}
\label{abc_bto_backward}
\end{subfigure}
\hspace{4mm}
\caption{Qualitative comparison of post hoc FTLE visualizations generated using basis flows for the ABC data set.}
\label{fig:abc_ftle_visualizations}
\end{figure}

\begin{table*}[!h]
\centering
\scalebox{0.8}{
\begin{tabular}{||c|c||c|c|b||c|c|c|c|d||}
\hline
\textbf{Interval} & \textbf{Reduction} & \textbf{L-BTO~(s)} & \textbf{L-MPI~(s)} & \textbf{Speed-up} & \textbf{Discarded \%} & \textbf{Greatest Max} & \textbf{Average Max} & \textbf{Total Average} & \textbf{Accuracy \%} \\
 &  &  &  & & & \textbf{L2-norm} & \textbf{L2-norm} & \textbf{L2-norm} & \\
\hline
5 & 1:1 & 0.0062 & 0.0089 & \textbf{1.4x} & 0.7 & 5.66$\times10^{-4}$ & 1.73$\times10^{-4}$  &  1.89$\times10^{-5}$ & \textbf{99.7} \\
\hline
10 & 1:1 & 0.0035 & 0.0089 & \textbf{2.5x} & 2.7 & 3.12$\times10^{-3}$  &  1.20$\times10^{-3}$ &  1.45$\times10^{-4}$ & \textbf{98.1} \\
\hline
5 & 1:8 & 0.0024 & 0.0044 & \textbf{1.8x} & 0.8 & 8.64$\times10^{-4}$  & 2.69$\times10^{-4}$  &  3.77$\times10^{-5}$ & \textbf{99.5} \\
\hline
10 & 1:8 & 0.0022 & 0.0059 & \textbf{2.6x} & 2.1 & 4.41$\times10^{-3}$  &  1.45$\times10^{-3}$  & 2.06$\times10^{-4}$ & \textbf{97.3} \\
\hline
5 & 1:27 & 0.0031 & 0.0038 & \textbf{1.2x} & 0.7 & 8.47$\times10^{-4}$  & 3$\times10^{-4}$  & 4.32$\times10^{-5}$ & \textbf{99.4} \\
\hline
10 & 1:27 & 0.0024 & 0.0094 & \textbf{3.9x} & 1.9 & 4.27$\times10^{-3}$ & 1.72$\times10^{-3}$  & 2.48$\times10^{-4}$ & \textbf{96.8} \\
\hline
5 & 1:64 & 0.0025 & 0.0037 & \textbf{1.4x} & 0.5 & 9.19$\times10^{-4}$ & 3.30$\times10^{-4}$  & 4.34$\times10^{-5}$ & \textbf{99.4} \\
\hline
10 & 1:64 & 0.0029 & 0.0039 & \textbf{1.3x} & 1.5 & 4.89$\times10^{-3}$  & 1.96$\times10^{-3}$  & 2.80$\times10^{-4}$ & \textbf{96.4} \\
\hline
\end{tabular}
}
\caption{Speed-up and reconstruction accuracy results for the Jet flow data set \textbf{WF2} experiments. Lagrangian-BTO and Lagrangian-MPI columns show average time per step in seconds. Reported results are measured across all intervals.}
\label{jet4_table}
\end{table*}

\begin{table*}[!h]
\centering
\scalebox{0.8}{
\begin{tabular}{||c|c||c|c|b||c|c|c|c|d||}
\hline
\textbf{Interval} & \textbf{Reduction} & \textbf{L-BTO~(s)} & \textbf{L-MPI~(s)} & \textbf{Speed-up} & \textbf{Discarded \%} & \textbf{Greatest Max} & \textbf{Average Max} & \textbf{Total Average} & \textbf{Accuracy \%} \\
 &  &  &  & & & \textbf{L2-norm} & \textbf{L2-norm} & \textbf{L2-norm} & \\
\hline
10 & 1:1 & 0.0027 & 0.0067 & \textbf{2.4x} & 4.1 & 5.06$\times10^{-4}$ & 1.52$\times10^{-4}$  & 3.93$\times10^{-5}$ & \textbf{99.5} \\
\hline
20 & 1:1 & 0.0026 & 0.0067 & \textbf{2.5x} & 8.4 & 8.19$\times10^{-4}$ & 4.83$\times10^{-4}$  & 1.52$\times10^{-4}$ & \textbf{98.0} \\
\hline
40 & 1:1 & 0.0026 & 0.0139 & \textbf{5.2x} & 15.9 & 3.97$\times10^{-3}$ & 2.25$\times10^{-3}$  & 9.02$\times10^{-4}$ & \textbf{88.5} \\
\hline
50 & 1:1 & 0.0025 & 0.0079 & \textbf{3.1x} & 19.3 & 6.04$\times10^{-3}$ & 3.59$\times10^{-3}$  & 1.64$\times10^{-3}$ & \textbf{79.1} \\
\hline
10 & 1:8 & 0.0024 & 0.0045 & \textbf{1.8x} & 3.2 & 9.68$\times10^{-4}$ & 1.67$\times10^{-4}$  & 3.96$\times10^{-5}$ & \textbf{99.4} \\
\hline
20 & 1:8 & 0.0023 & 0.0107 & \textbf{4.6x} & 7.9 & 1.74$\times10^{-3}$  & 7.05$\times10^{4}$  & 2.19$\times10^{-4}$ & \textbf{97.2} \\
\hline
40 & 1:8 & 0.0027 & 0.0046 & \textbf{1.7x} & 15.5 & 6.19$\times10^{-3}$ &   2.92$\times10^{-3}$  & 1.18$\times10^{-3}$ & \textbf{85.6} \\
\hline
50 & 1:8 & 0.0030 & 0.0045 & \textbf{1.4x} & 18.9 & 7.74$\times10^{-3}$ & 4.32$\times10^{-3}$  & 2.02$\times10^{-3}$ & \textbf{74.3} \\
\hline
\end{tabular}
}
\caption{Speed-up and reconstruction accuracy results for the Nyx simulation data set \textbf{WF2} experiments. Lagrangian-BTO and Lagrangian-MPI columns show average time per step in seconds. Reported results are measured across all intervals.}
\label{nyx_table}
\end{table*}

%
Lagrangian-BTO demonstrates an average of 2.3x speed-up when compared to Lagrangian-MPI.
%
%
For the ABC data set, reconstruction accuracy does not significantly deteriorate when using a smaller number of particles.
However, for configurations with an interval equal to 100, a large number of particles exit the node boundaries and are terminated. 
For example, when the interval is set to 100 and a data reduction of 1:1 is used, over 1.7M particles are terminated every interval~(approximately 10\% of all seeds initially placed), which results in the reconstruction accuracy reducing to 95.5\%. 
For the ABC data set, the number of particles terminated per interval approximately doubles every time the interval doubles.
We observe that all configurations that have an interval of 25 or 50 achieve a reconstruction accuracy that is greater than 99\%.
When considering the average L2-norm error of individual intervals of the ABC data set in Figure~\ref{ABC_Bubble} and~\ref{ABC_Line}, we observe a sinusoidal behavior in the error that is due to the sinusoidal ABC analytical function.

Figures~\ref{abc_mpi_backward} and~\ref{abc_bto_backward} compare FTLE visualizations generated using approximately 2M basis flows computed over 100 time steps and saved by each technique (configuration in row 6 of Table~\ref{abc_table}).
Although we observe small discontinuities in the ridges of the FTLE field generated using Lagrangian-BTO basis flows, the overall quality of the FTLE visualization is similar.
%
%

\subsection{Jet Flow Simulation}
\label{sec:jet4}
The Jet data set is a simulation of a jet of high-velocity fluid entering a medium at rest.
It was created using the Gerris Flow Solver~\cite{popinet2003gerris}.
The vector field is defined over a $128\times256\times128$ uniform grid and we use a total of 500 time steps (previously subsampled).
%
%
These \textbf{WF2} experiments tested the performance of the Lagrangian-BTO when reconstructing this turbulent data set by considering two intervals~(5, 10) and four data reduction values~(1:1, 1:8, 1:27, 1:64). 
All tests used 16 nodes, 64 MPI tasks, with 4 MPI tasks using 4 GPUs on each node. 
Table~\ref{jet4_table} contains configuration information and results of the Jet data set.
Lagrangian-BTO demonstrates an average speed-up of 2x when compared to Lagrangian-MPI for this data set.
The domain contains regions of high velocity resulting in the reconstruction accuracy of the Lagrangian-BTO analysis filter being adversely affected as the interval increases.
For configurations with the interval set to 5, the data reduction is less consequential, and all configurations achieve greater than 99\% accuracy. 
However, for configurations with the interval set to 10, the reconstruction accuracy decreases from 98.1\% to 96.4\% as data reduction ranges from 1:1 to 1:64.
Thus, similar to the result observed in the strong scaling study in Section~\ref{sec:clover}, the combination of larger interval and reduced number of particles results in less accurate reconstruction. 

\begin{figure}[!t]
\begin{subfigure}{0.45\linewidth}
\centering
\includegraphics[width=1\linewidth]{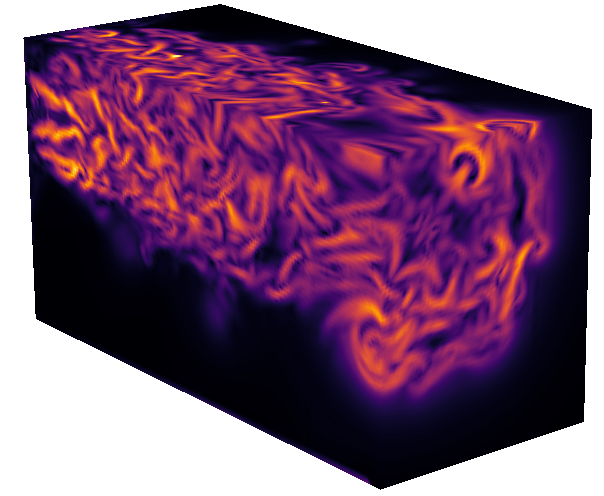}
\caption{Lagrangian-MPI}
\label{jet_mpi_forward}
\end{subfigure}
\begin{subfigure}{0.55\linewidth}
\centering
\includegraphics[width=1\linewidth]{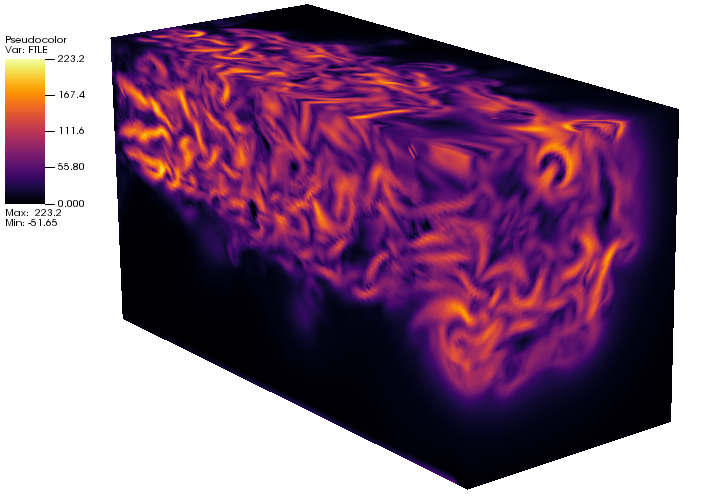}
\caption{Lagrangian-BTO}
\label{jet_bto_forward}
\end{subfigure}
\caption{Qualitative comparison of post hoc FTLE visualizations generated using basis flows for the Jet flow data set. 
}
\label{fig:jet_ftle_visualizations}
\end{figure}

Figures~\ref{jet_mpi_forward} and~\ref{jet_bto_forward} compare visualizations produced using 4.2M basis flows saved by each method (configuration in row 2 of Table~\ref{jet4_table}).
The Lagrangian-BTO FTLE visualization has some instances of a less pronounced structure in the FTLE ridges.
However, the differences are localized and relatively small compared to the overall information conveyed by the images.

\begin{figure}[!t]
\hspace{5mm}
\begin{subfigure}{0.38\linewidth}
\centering
\includegraphics[width=1\linewidth]{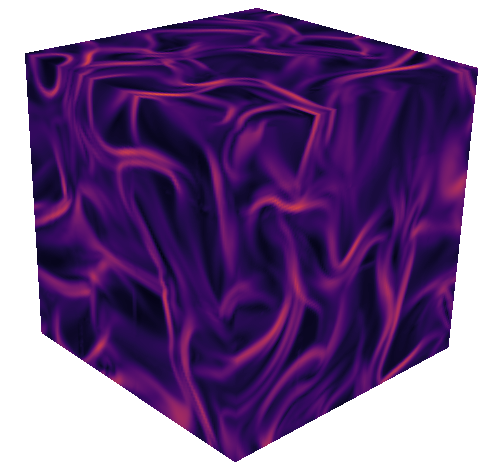}
\caption{Lagrangian-MPI}
\label{nyx_mpi_backward}
\end{subfigure}
\hspace{2mm}
\begin{subfigure}{0.45\linewidth}
\centering
\includegraphics[width=1\linewidth]{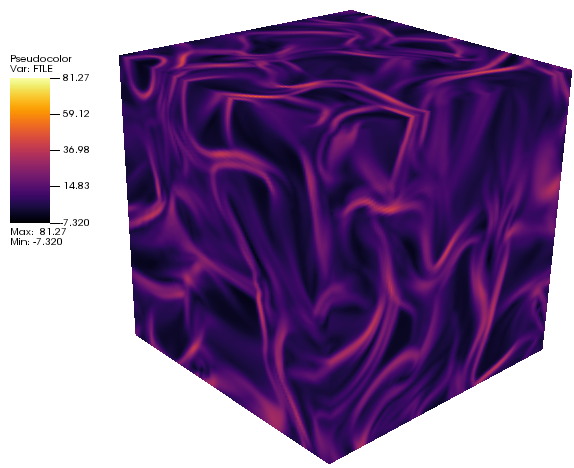}
\caption{Lagrangian-BTO}
\label{nyx_bto_backward}
\end{subfigure}
\caption{Qualitative comparison of post hoc FTLE visualizations generated using basis flows for the Nyx data set.}
\label{fig:nyx_ftle_visualizations}
\end{figure}

\subsection{Nyx Cosmology Simulation} 
\label{sec:nyx}
The Nyx simulation is a N-body and gas dynamics code for large-scale cosmological simulations~\cite{almgren2013nyx}. 
We use a Nyx test executable named TurbForce to generate 500 cycles of a $128^3$ data set with an average time step of $0.002$.
The generated flow field grows in turbulence over the duration of the simulation.
These \textbf{WF2} experiments with the Nyx data set test the reconstruction accuracy of Lagrangian-BTO when considering four intervals (10, 20, 40, 50) and two data reduction values (1:1, 1:8).
Similar to previous setups, we use 64 MPI tasks evenly distributed across 16 nodes with each MPI task using a single GPU.
Table~\ref{nyx_table} details test configurations and results for the Nyx data set.

Lagrangian-BTO demonstrates an average speed-up of 2.8x over Lagrangian-MPI.
Configurations with an interval of 10 or 20 have high reconstruction accuracies greater than 97\%. 
However, as the interval increases, the reconstruction accuracy is adversely affected by the large number of particles terminated and the significant turbulence in this data set.
Further, Lagrangian-BTO can reconstruct the field relatively more accurately when using a 1:1 data reduction compared to a 1:8 configuration.
Figures~\ref{nyx_mpi_backward} and~\ref{nyx_bto_backward} compare FTLE visualizations generated using approximately 2M basis flows saved by each technique (configuration in row 1 of Table~\ref{nyx_table}).
We observe qualitatively comparable FTLE ridge structures. 

\section{Conclusion and Future Work}

Performing accurate exploratory analysis and visualization of time-varying vector fields is particularly challenging in sparse temporal settings.
Our proposed algorithm represents an important step in expanding
the Lagrangian in situ reduction and post hoc exploration paradigm to be viable
for large-scale simulations.
Predecessor work demonstrated that accuracy-storage tradeoffs were clearly
superior to the traditional approach~\cite{agranovsky2014improved}.
However, this work did not place a significant emphasis on minimizing
in situ execution times and ensuring scalability.
Our work addresses this point, and the corresponding reduced execution times (2x to 4x)
and improved scalability remove another barrier to adoption.

We feel the most surprising result from our work is the rate at which we achieve high reconstruction accuracy.
Clearly, terminating particles at block boundaries makes
for less useful basis flows,
and can create issues where post hoc exploration accuracy suffers.
However, our results empirically show that this feared case happens relatively infrequently
in practice and is limited to instances of long intervals, i.e., large integration times.
Our proposed approach works particularly well for practical configurations with a short or medium interval, while delivering the same performance benefits.
31 of our 35 tests gave accuracy that was greater than 96\%. 
In addition to our quantitative evaluation, we qualitatively showed that comparable FTLE visualizations can be generated using basis flows extracted at a fraction of the in situ cost.
These evaluations are relative to the Agranovsky et al. approach,
which incurred significantly more cost in execution time and scales poorly.
Further, while we are slightly less accurate than Agranovsky et al., our approach would still be much more accurate than the traditional Eulerian approach for time-varying flow visualization in sparse temporal settings.

In terms of future work, we aim to store particle trajectory termination locations on the boundary, develop Lagrangian-based advection schemes that can consume flow maps with trajectories stopping (and starting) at arbitrary times, integrate adaptive variable duration variable placement~(VDVP) techniques~\cite{sane2019interpolation}, and consider the use of flow field characteristics to guide flow map extraction.
Further, research is required to address edges cases and the limitations of a communication-free model demonstrated by our experiments.
We believe our work is foundational because it evaluated the use of a simple communication-free model and demonstrated improved scalability with only a reasonable loss of accuracy.
%
%
Remaining communication-free is particularly essential for Lagrangian analysis techniques that aim to scale well and remain within in situ constraints of large-scale simulations.

\acknowledgments{This research was supported by the Exascale Computing Project
(17-SC-20-SC), a collaborative effort of the U.S. Department of
Energy Office of Science and the National Nuclear Security Administration.}

\bibliographystyle{abbrv-doi}

\bibliography{ms}
\end{document}